\documentclass[lettersize,journal]{IEEEtran}
\usepackage{amsthm}
\usepackage{amsmath}
\usepackage{times}
\usepackage{graphicx}
\usepackage[colorlinks=true, allcolors=blue]{hyperref}
\usepackage{booktabs}
\usepackage{array}
\usepackage{makecell}
\usepackage{pifont}
\usepackage{caption}
\usepackage{authblk}
\usepackage[T1]{fontenc}
\usepackage{multirow}
\usepackage{hyperref}
\usepackage{longtable}
\usepackage{makecell}
\setcellgapes{4pt}
\usepackage{tikz}
\usetikzlibrary{shapes, positioning, arrows.meta}
\makegapedcells

\usepackage[english]{babel}

\usepackage{amssymb}

\usepackage{array}
\usepackage{booktabs}
\usepackage{multirow}
\usepackage{ragged2e}
\newcolumntype{P}[1]{>{\RaggedRight\hspace{0pt}}p{#1}}
\usepackage[final]{microtype}
\usepackage{enumitem}
\setlist{nosep,leftmargin=*}

\newenvironment{tightmath}{%
  \setlength{\abovedisplayskip}{6pt}%
  \setlength{\belowdisplayskip}{6pt}%
  \setlength{\abovedisplayshortskip}{4pt}%
  \setlength{\belowdisplayshortskip}{4pt}%
  \setlength{\jot}{2pt}%
}{}

\AtBeginDocument{%
  \fontdimen2\font=0.18em 
  \fontdimen3\font=0.1em  
  \fontdimen4\font=0.06em  
}

\begin{document}
%
\title{\color{black}\textsc{PeerSync}: Accelerating Containerized Model Inference at the Network Edge}
%
%
%

\author{Yinuo~Deng,
        Hailiang~Zhao,~\IEEEmembership{Member,~IEEE},
        Dongjing~Wang,
        Peng~Chen,
        Wenzhuo~Qian,
        Jianwei~Yin,
        Schahram~Dustdar,~\IEEEmembership{Fellow,~IEEE},~and
        Shuiguang~Deng,~\IEEEmembership{Senior~Member,~IEEE}
\thanks{
    This work was supported in part by the National Science Foundation of China (62125206, 62502441, 62202131), and Zhejiang Provincial Natural Science Foundation of China (LD25F020002, LZ25F020010). Hailiang Zhao's work was supported in part by the Zhejiang University Education Foundation Qizhen Scholar Foundation.
}
\thanks{
    Yinuo Deng, Peng Chen, Wenzhuo Qian, Jianwei Yin, and Shuiguang Deng are with the College of Computer Science and Technology, Zhejiang University. Emails: \{yinuo, pgchen, qwz, zjuyjw, dengsg\}@zju.edu.cn.
}
\thanks{
    Hailiang Zhao is with the School of Software Technology, Zhejiang University. Email: hliangzhao@zju.edu.cn.
}
\thanks{
    Dongjing Wang is with the College of Computer Science and Technology, Hangzhou Dianzi University. Email: dongjing.wang@hdu.edu.cn.
}
\thanks{Schahram Dustdar is with the Distributed Systems Group at the TU Wien and with ICREA at the UPF, Barcelona. Email: dustdar@dsg.tuwien.ac.at. }
\thanks{Hailiang Zhao and Shuiguang Deng are corresponding authors.}
}

\IEEEaftertitletext{\vspace{-2\baselineskip}}
\maketitle

\begin{abstract}
    Efficient container image distribution is crucial for enabling machine learning inference at the network edge, where resource limitations and dynamic network conditions create significant challenges. In this paper, we present \textsc{PeerSync}, a decentralized P2P-based system designed to optimize image distribution in edge environments. \textsc{PeerSync} employs a popularity- and network-aware download engine that dynamically adapts to content popularity and real-time network conditions. \textsc{PeerSync} further integrates automated tracker election for rapid peer discovery and dynamic cache management for efficient storage utilization. We implement \textsc{PeerSync} with 8000+ lines of Rust code and test its performance extensively on both large-scale Docker-based emulations and physical edge devices. Experimental results show that \textsc{PeerSync} delivers a remarkable speed increase of 2.72$\times$, 1.79$\times$, and 1.28$\times$ compared to the Baseline solution, Dragonfly, and Kraken, respectively, while significantly reducing cross-network traffic by 90.72\% under congested and varying network conditions.
\end{abstract}


%
\IEEEpeerreviewmaketitle

\section{Introduction}\label{s1}
Model inference at the network edge is becoming increasingly popular in applications such as edge-based video analytics, real-time inference in smart cities, and IoT-based predictive maintenance. To support these scenarios, containerization technologies, such as Docker, are widely used to enable the rapid deployment and management of workloads at scale \cite{8666479,273798}. Container images, representing multi-layered filesystems, are not merely software artifacts but structured data that must be efficiently stored, distributed, and retrieved across diverse environments, ranging from cloud platforms to the network edge \cite{costa2022orchestration,253358,253378,8567693,d7o,kangjin2017fid}.

Deploying a containerized application involves retrieving the necessary image layers from a container registry (e.g., Docker Hub), unpacking them, mounting them using a layered filesystem, and initiating the container entrypoint. These images, structured as a stack of layers representing filesystem segments, are stored and managed in centralized container registries. While these registries efficiently facilitate container image distribution in cloud environments, they encounter unique challenges in edge computing scenarios: (i) \textit{High latency:} Distributed edge devices face substantial delays when fetching image layers over wide-area networks, leading to degraded responsiveness and increased {cold-startup} times for services \cite{253358,becker2021edgepier,278314}; (ii) \textit{Inefficiency under high concurrency:} The non-parallel architecture of container registry uplinks often causes severe bottlenecks when handling concurrent image-pulling requests from multiple edge devices \cite{278314}. The growing size of AI models exacerbates this issue. 
To address these limitations, peer-to-peer (P2P) architectures have been explored to accelerate container image distribution while alleviating the load on centralized registries \cite{253358,273798,8814532,194430,254400,zheng2018wharf,kangjin2017fid,liang2016hdid,d7o,kraken}. P2P systems inherently offer distributed replication, parallel data transfers, and resilience against single points of failure. Although P2P-based solutions have shown promise in cloud environments, their effectiveness in edge settings is constrained by the unique characteristics of edge computing:
\begin{itemize}
    \item \textit{Limited network bandwidth.} Unlike high-capacity data center networks, edge environments often has significantly constrained bandwidth, hindering fast data exchange \cite{costa2022orchestration}.
    \item \textit{Dynamic network topology.} The dynamic nature of edge networks introduces frequent connectivity changes, requiring robust mechanisms to adaptively manage edge devices and maintain consistency without relying on centralized components such as static trackers \cite{jia2008designs}.
    \item \textit{Low storage scalability.} Portable edge devices, which typically rely on constrained storage media like SD cards or embedded flash memory, often have limited storage. Unlike cloud data centers equipped with technologies like Storage Area Networks \cite{tate2018introduction}, their scalability is inherently restricted. Over time, the accumulation of large AI-related images and other data, such as raw sensor outputs, can quickly deplete available storage space.
\end{itemize}

\begin{table*}[!ht]
    \scriptsize
    \renewcommand{\arraystretch}{0.05}
    \setlength{\aboverulesep}{0pt}
    \setlength{\belowrulesep}{0pt}
    \setlength{\abovetopsep}{0pt}
    \setlength{\belowbottomsep}{0pt}
    \setlength{\cmidrulesep}{0pt}
    \setlength{\tabcolsep}{3pt}         
    \setlength{\tabcolsep}{3pt}
    \caption{Comparison between \textsc{PeerSync} and state-of-the-art works. `F.D.': `Fully Decentralized', `S.D.': `Semi Decentralized'.}
    \centering
    \label{table:comparison}
    \begin{tabular}{c|cccccc}
    \toprule
    {Work} & \textsc{PeerSync} & {Kraken} \cite{kraken} & {Dragonfly} \cite{d7o} & {Starlight} \cite{278314} & {EdgePier} \cite{becker2021edgepier} & {FID} \cite{kangjin2017fid}  \\
    \midrule
    {Architecture} & P2P (F.D.) & P2P (S.D.) & P2P (S.D.) & C/S & P2P (F.D.) & P2P (S.D.) \\
    {P2P flavor} & Impl. from scratch & Impl. from scratch & Impl. from scratch & - & IPFS & BitTorrent \\
    {Scenario} & General & Cloud & Cloud & WAN & Edge & General \\
    {Tracker placement} & dynamic & static & static & - & static & static \\
    \bottomrule
    \end{tabular}
\end{table*}

While P2P-based solutions such as EdgePier \cite{becker2021edgepier}, Kraken \cite{kraken}, and Starlight \cite{278314} address some of these issues and achieve great performance in most cases, they are still based on the traditional P2P architecture and consequently fall short in effectively balancing network efficiency, storage optimization, and adaptability to edge-specific conditions. More specifically, they lack fine-grained control mechanisms and rely on static trackers, making them unsuitable for fluctuating network conditions. Starlight \cite{278314} minimizes container startup latency by redesigning deployment protocols and storage formats but requires intrusive modifications to cloud registries, limiting its applicability. Moreover, existing solutions often neglect factors such as dynamic network quality, content popularity, and effective storage utilization, leaving room for improvement. For instance, our preliminary experiments (detailed in Sec. \ref{s3}) demonstrate that maintaining a single copy of any image layer within a local area network (LAN) can significantly enhance image-fetching speed without overloading local storage.

\textcolor{black}{Crucially, the edge is not a minimized data center. Systems like Kraken \cite{kraken} and Dragonfly \cite{d7o}, though highly optimized for homogeneous, stable cloud networks, assume persistent centralized trackers or super-nodes that become single points of failure under edge churn. EdgePier \cite{becker2021edgepier} improves decentralization but still presumes stable anchor nodes, which rarely exist in mobile or intermittently connected edge deployments. Meanwhile, proposals leveraging IPFS or BitTorrent \cite{kangjin2017fid} inherit protocol overheads ill-suited for low-resource devices. Even more recent learning-based approaches \cite{mohajer2024dynamic,8818442,8818403,9154603}, while theoretically appealing, require GPU/NPU resources, extensive training data, and stable feedback loops, rendering them impractical on typical edge hardware such as Raspberry Pi.} To overcome these challenges, we propose \textsc{PeerSync}, a non-intrusive, P2P-based system tailored to the unique demands of the edge. PeerSync constructs a fully decentralized P2P network across different LANs. It features a P2P image download engine that optimizes distribution by periodically calculating the required content pieces for each missing layer of a requested image, leveraging a scoring system that evaluates both content popularity and the quality of the network connection between peers and the requesting edge device. Unlike traditional P2P architectures that rely on manually configured trackers, PeerSync autonomously elects trackers based on real-time metrics, including bandwidth availability and resource utilization, ensuring resilience and performance. PeerSync also employs a selective deletion mechanism to optimize storage utilization, maintaining only essential image layers within each LAN. We compare PeerSync with state-of-the-art solutions in Table \ref{table:comparison}. In summary, our main contributions are as follows:
\begin{enumerate}
    \color{black}
    \item We design an edge container image distribution system \textsc{PeerSync}, featuring a high-performance downloading engine, fault-tolerant tracker with self-healing ability, and collaborative dynamic space reclamation for efficient caching.
    \item \textsc{PeerSync} dynamically assigns scores to peers. Its scoring function jointly optimizes for (i) local network proximity (e.g., same subnet), (ii) real-time bandwidth stability, and (iii) layer popularity. This ensures that the majority of traffic remains within the LAN, directly addressing the high-latency and limited-bandwidth constraints of WAN links.
    \item We implement PeerSync with 8000+ lines of Rust code, resulting in a statically linked binary of just 8.8 MB, making it easily deployable on edge devices. \textsc{PeerSync} is compatible with the Open Container Initiative (OCI) standard and integrates transparently with existing container runtimes.
    \item We conducted extensive experiments to verify \textsc{PeerSync}'s performance under different network conditions. On average, \textsc{PeerSync} achieves 2.72$\times$ faster distribution than the Baseline, 1.79$\times$ faster than Dragonfly, and 1.28$\times$ faster than Kraken. Additionally, \textsc{PeerSync} reduces peak cross-network traffic by 90.72\% under congested and variable network conditions.
\end{enumerate}

\section{Motivation}\label{s3}
In this section, we empirically and analytically dissect the limitations of two representative approaches: Kraken \cite{kraken} and Starlight \cite{278314}, motivating the need for a more efficient solution tailored to edge computing environments.

\subsection{Traditional P2P Approaches}\label{s3.1}
P2P architectures such as Napster \cite{carlsson2001rise}, BitTorrent \cite{bittorrent}, and IPFS \cite{benet2014ipfs} enable decentralized content sharing by allowing nodes to act as both consumers and providers. While these systems excel in cloud or data center settings with abundant bandwidth and stable connectivity, they suffer from a critical flaw in edge environments: \textit{the lack of bandwidth awareness and network topology sensitivity}. This leads to inefficient utilization of scarce inter-LAN links and undermines the very benefits P2P promises.

\begin{figure}[t]
    \centering
    \includegraphics[width=3.2in]{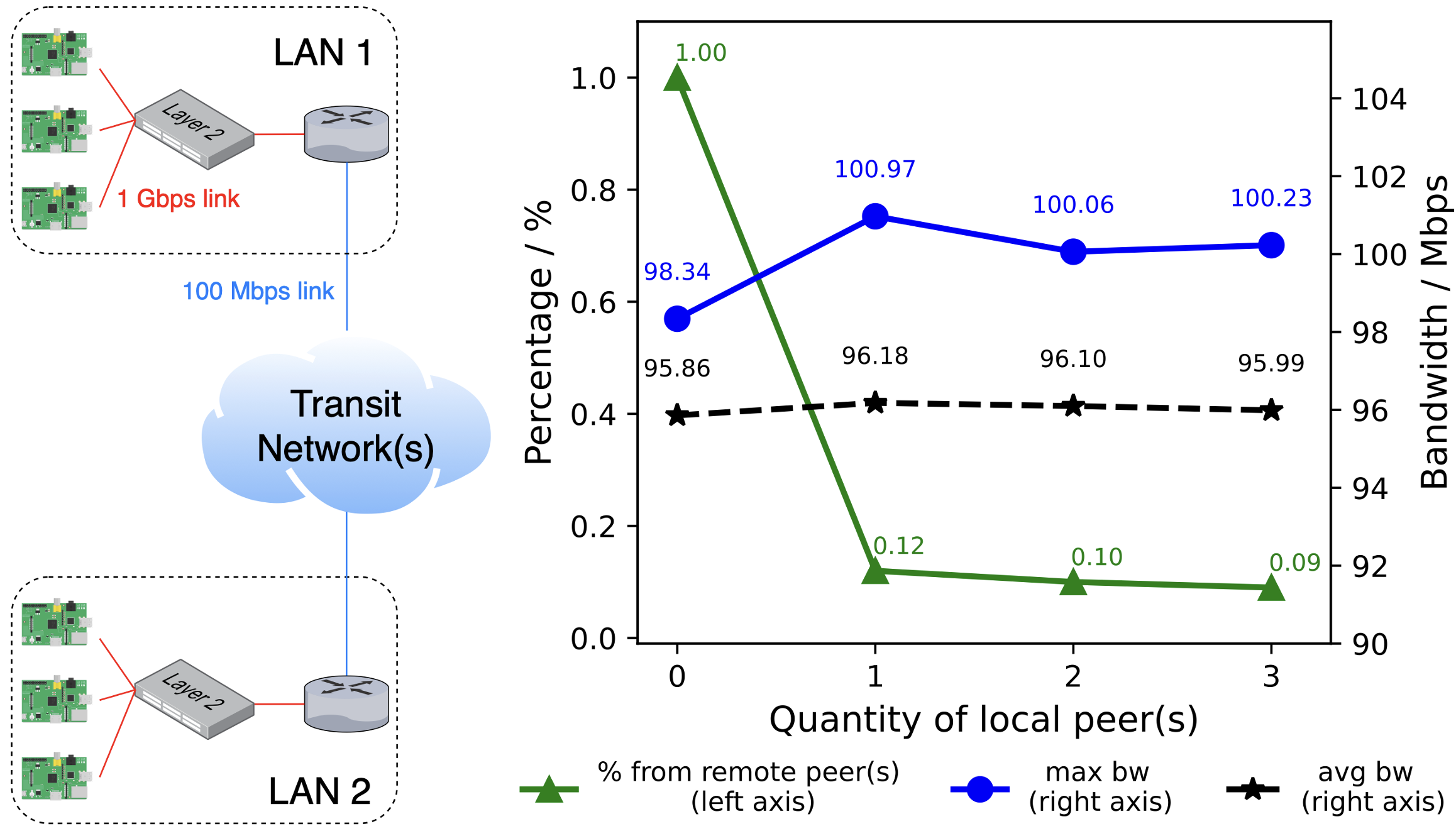}
    \caption{An edge computing environment with 2 LANs and the corresponding observed results when using BitTorrent for image downloading.}
    \label{fig_motivation}
\end{figure}

\textcolor{black}{To illustrate this, we conducted an experiment using Kraken \cite{kraken}, which builds upon BitTorrent's protocol. As shown on the left side of Fig.~\ref{fig_motivation}, our testbed comprised two LANs connected via a 100 Mbps link, representative of typical edge-to-edge or edge-to-cloud uplinks. Each LAN included three Raspberry Pi 4 Model B devices (each with 4 GB RAM and 32 GB eMMC storage) connected through a 1 Gbps switch. In LAN 1, two hosts served as seeders for a large container image; download requests originated from LAN 2, with 1, 2, and 3 local peers progressively added as potential sources. The results (right side of Fig.~\ref{fig_motivation}) reveal a striking inefficiency: even when local peers were available in LAN 2, Kraken still fetched approximately 10\% of the image blocks from remote peers in LAN 1. Crucially, this small fraction consumed over 95\% of the inter-LAN bandwidth. This behavior stems from BitTorrent's peer selection strategy, which prioritizes global peer diversity and piece availability over network locality. As a result, the narrow uplink becomes saturated, delaying downloads for all nodes and negating the latency advantage of local caching. Beyond bandwidth waste, P2P systems like Kraken impose significant storage pressure on edge devices. In a separate experiment, we distributed the top-10 most popular Docker Hub images across Kraken nodes. After downloading and decompressing these images, each node consumed 1408.54 MiB of disk space. For resource-constrained edge devices (e.g., those using SD cards or embedded flash with limited write endurance), such storage overhead is unsustainable, especially when images accumulate over time alongside sensor data or model checkpoints.}


Furthermore, Kraken's reliance on a centralized tracker for peer discovery introduces a single point of failure. In dynamic edge environments, where devices frequently join, leave, or experience intermittent connectivity, tracker outages lead to network fragmentation, prolonged peer discovery latency, and often require manual intervention to restore service \cite{bep3, neglia2007availability, kononova2021lifecycle}. This centralized coordination model is fundamentally at odds with the autonomy and resilience required in edge deployments.

\subsection{Image Structure-based Approaches}\label{s3.2}

\begin{figure}[t]
\centering
    \includegraphics[width=3.3in]{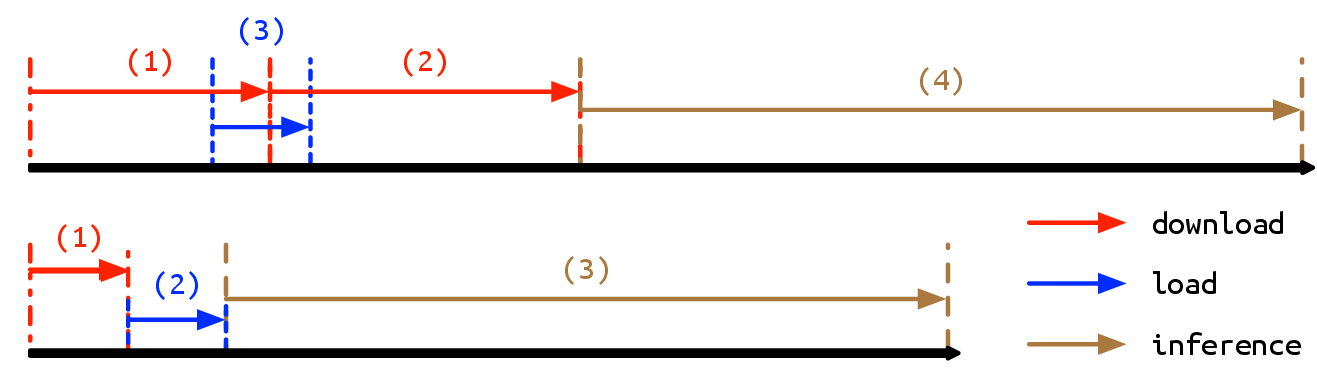} 	
    \caption{Timeline diagram of a large model inference task. \textit{Top: Lazy Loading (e.g., Starlight \cite{278314}):} Download components (PyTorch, cuDNN library, etc.) $\to$ Download the model $\to$ Load the ML framework $\to$ Perform inference. \textit{Bottom: P2P (e.g., \textsc{PeerSync}):} Download the full image $\to$ Load the ML framework $\to$ Perform inference.}
    \label{fig_inf_starlight_ps}
\end{figure}

\textcolor{black}{
An alternative strategy, exemplified by Starlight \cite{278314}, seeks to minimize container startup latency by analyzing image structures and runtime behavior. Starlight removes unnecessary components (a process known as debloating \cite{rastogi2017cimplifier, slim}) and decouples the download and execution phases through lazy loading: it prioritizes ``hot'' files (e.g., libraries) based on runtime traces collected during a profiling run. While effective for certain applications, this approach faces three critical limitations in edge AI scenarios: 
\begin{itemize}
    \item \textit{Starlight assumes workload homogeneity and predictability}. Starlight requires users to convert images into a custom format and execute them once to collect file-access traces. This profiling step assumes that future executions will follow similar patterns, which breaks down when deploying diverse or evolving AI workloads (e.g., switching from object detection to LLM inference). At scale, maintaining accurate traces for hundreds of edge applications becomes impractical.
    \item \textit{Starlight offers minimal benefit for large-model inference tasks}. As shown in Fig.~\ref{fig_inf_starlight_ps}, inference pipelines are typically dominated by two phases: (1) fetching the model and framework, and (2) computation. Crucially, the entire model must be available before inference can begin, as models are often stored as a few large files (e.g., in Safetensors \cite{safetensors} or PyTorch format \cite{paszke2019pytorch}). Consider Meta's Llama 3.1 \cite{dubey2024llama} (\texttt{ai/meta-llama:3.1-8B-Instruct-cuda-12.6}). When uncompressed, this image occupies approximately 21.4 GiB of disk space, composed of only two major parts: \textit{Model weights}: 15 GiB (70.10\%), stored in just 4 files using Safetensors \cite{safetensors}; \textit{ML framework and dependencies}: 6.2 GiB (28.97\%), including PyTorch \cite{paszke2019pytorch} and CUDA libraries. In such cases, downloading libraries first provides negligible speedup since the critical path remains the transfer of the 15 GiB model file. Trace-based methods like Starlight cannot accelerate this bottleneck because they optimize \textit{what} and \textit{when} to download, not \textit{how fast} the data can be fetched.
\end{itemize}
}

 In addition, Starlight remains dependent on centralized registries. Despite its optimizations, it still pulls base images from Docker Hub or private registries, inheriting all the latency, bandwidth, and availability challenges of cloud-centric architectures. In disconnected or bandwidth-constrained edge settings, this dependency severely limits deployability.

{
\color{black}
\subsection{Insights}\label{s3.3}
Both Kraken and Starlight illustrate shortcomings in existing approaches. Kraken, while leveraging P2P architectures, suffers from bandwidth inefficiencies and excessive reliance on centralized trackers. Starlight, despite its innovative lazy-loading approach, struggles with scalability across diverse model inference workloads and fails to address large model file distribution effectively. These limitations highlight the need for a decentralized, bandwidth-aware solution that maximizes local resource utilization, minimizes bandwidth waste, and adapts dynamically to the constraints of edge environments.
}

\section{The \textsc{PeerSync} System}\label{s4}

\subsection{Problem Formulation}\label{s4.1}
\textcolor{black}{We consider a distributed environment where nodes (including cloud instances and edge devices) pull container images from a central registry. Each image consists of a manifest and multiple immutable layers. When many nodes concurrently pull the same image, the registry becomes a bottleneck, leading to high latency and WAN bandwidth consumption. Let $\mathcal{P}$ be the set of peers, each holding a subset of layers $\mathcal{L}_p \subset \mathcal{L}$, where $p \in \mathcal{P}$. Peers are connected via heterogeneous, dynamic network links with unknown bandwidth and availability. A requesting peer $p_\text{req}$ must fetch all layers of a target image as quickly as possible, without prior knowledge of which peers have which layers or their current upload capacity. The problem is to minimize the end-to-end image pull time by opportunistically leveraging P2P transfers among peers, while operating without global coordination, runtime modifications, or assumptions about peer stability or connectivity.}




\subsection{System Architecture Overview}\label{s4.2}

\begin{figure}[t]
    \includegraphics[width=3.45in]{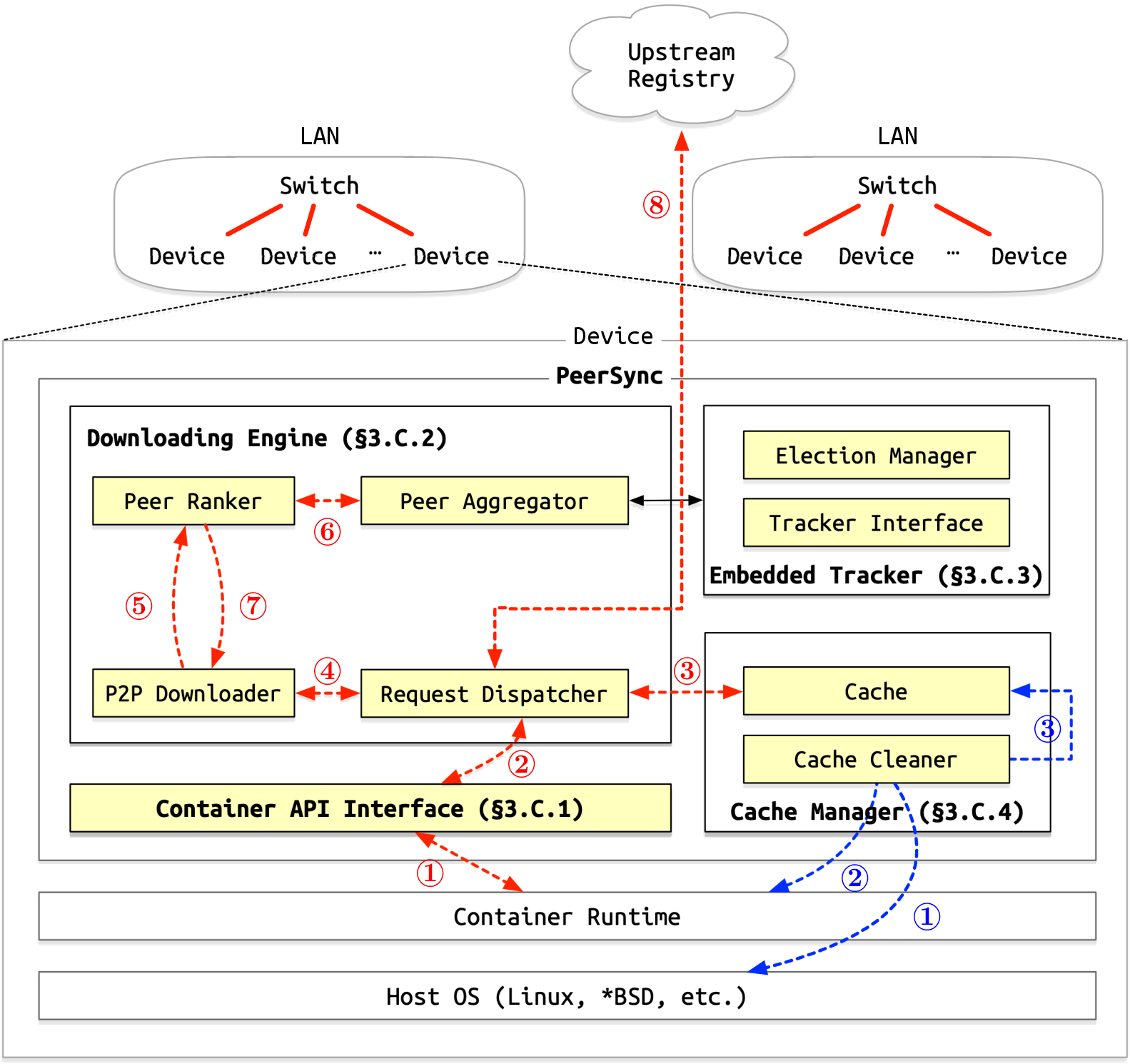} 	
    \caption{The architecture of \textsc{PeerSync}.}
    \label{fig_overview}
\end{figure}

\textsc{PeerSync} is designed as a drop-in enhancement for existing container ecosystems. As illustrated in Fig.~\ref{fig_overview}, the system comprises four core components: \emph{API Interface}, \emph{Downloading Engine}, \emph{Embedded Tracker}, and \emph{Cache Manager}, which are orchestrated along two orthogonal threads: (i) the pull path, which handles on-demand layer retrieval, and (ii) the cache maintenance path, which runs continuously in the background to manage storage and peer metadata. \textcolor{black}{The entire system exposes a standard OCI Distribution API, making it fully compatible with Docker, Podman, and other OCI-compliant clients. No modifications to the container runtime or user workflow are required; peers simply configure their client to use \textsc{PeerSync} as the registry endpoint.}

A typical pull request begins when a container runtime issues an HTTP GET for a layer. The request is intercepted by the \emph{API Interface}, which checks local cache availability. If the layer is absent, the request is forwarded to the \emph{Downloading Engine}, where a lightweight \emph{Request Dispatcher} decides whether to fetch the layer from upstream (via HTTP) or initiate a P2P transfer based on multiple factors. When P2P is selected, the engine queries the \emph{Embedded Tracker} for candidate peers and coordinates block-level downloads. Upon successful retrieval, the layer is stored in the local \emph{Cache}, and the response is returned to the client. Concurrently, the \emph{Cache Manager} monitors global access patterns across layers and performs eviction based on a cost-aware policy that considers layer size, replication count within the LAN, and reuse likelihood. The \emph{Embedded Tracker}, operating independently, maintains an up-to-date view of nearby peers through periodic gossip and self-election, ensuring coordination remains decentralized and resilient.

\textcolor{black}{\textsc{PeerSync} attempts to join the P2P swarm during application startup. During bootstrap, \textsc{PeerSync} reads predefined nodes from the configuration and broadcasts bootstrap messages locally to join the swarm. If the bootstrap phase fails, \textsc{PeerSync} runs in a degraded mode as a proxy for the upstream registry and continues to bootstrap until a full P2P connection can be established. The predefined node list is dynamically updated with high-uptime peers, enabling \textsc{PeerSync} to withstand environmental changes.}

\subsection{Component Design} \label{s4.3}

\subsubsection{Container API Interface}\label{s431}
Modern container ecosystems achieve remarkable interoperability through adherence to the OCI standards. \textcolor{black}{The OCI defines both an \textit{Image Specification} (governing image format and manifest structure) and a \textit{Distribution Specification} (defining HTTP-based registry APIs for pulling images). To ensure \textsc{PeerSync} operates as a drop-in replacement without requiring any modification to existing runtimes or user workflows, it must fully comply with these standards. Crucially, the container runtime initiates image pulls via a well-defined sequence: one request for the image manifest (a JSON document listing layer digests and metadata), followed by one request per missing layer (each returning a gzipped tarball). Any deviation from this protocol would break compatibility. Thus, the primary motivation is to provide a transparent, standards-compliant facade that hides the internal P2P complexity while preserving the exact API contract expected by the runtime.}

The Container API Interface implements a subset of the OCI Distribution Specification. It serves as the sole external entry point for the container runtime.
\begin{itemize}
    \item \textit{Manifest Handling.} The manifest is typically small (often $<5$ KB) but frequently accessed and subject to upstream updates (e.g., \textsc{latest} tag). To ensure low-latency responses and immediate consistency, manifests are stored in an in-memory cache. This cache is kept up-to-date by periodically polling the upstream registry or by invalidating entries upon detecting a new pull request for a potentially stale tag.
    \item \textit{Layer Handling.} Layer requests are not served directly. Instead, the interface acts as a lightweight dispatcher, forwarding each blob request (identified by its digest) to the Downloading Engine (Sec.~\ref{s4.3}). The engine is responsible for fulfilling the request, whether from local cache, a LAN peer, or the upstream registry, and streaming the data back through this interface.
\end{itemize}

\begin{figure}[t]
    \centering
    \includegraphics[width=2in]{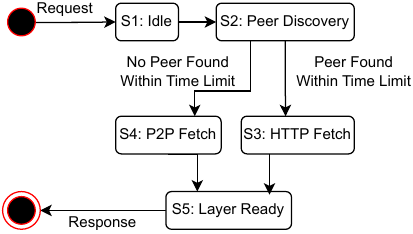}
    \caption{\textcolor{black}{The state machine of P2P downloading in \textsc{PeerSync}.}}
    \label{partial-p2p-state-machine}
\end{figure}

\subsubsection{Downloading Engine}\label{s432}
A naive P2P approach, which attempts to find peers for every layer, fails to account for the empirical reality of container image composition. As shown in Table~\ref{table:block_size_dist}, nearly half of all layers are smaller than 512 KiB, with a median size of just 1.03 MiB. For such small payloads, the latency of peer discovery in a Distributed Hash Table (DHT) or even a local multicast can exceed the time required to download the layer directly from a nearby registry. Conversely, large layers (e.g., base OS images) can benefit immensely from parallel, multi-source P2P downloads. The engine's core motivation is to make a per-layer, context-aware decision that optimizes for both speed and system stability, dynamically choosing between P2P and direct upstream fetching.

\begin{table}[h!]
    \centering
    \scriptsize
    \renewcommand{\arraystretch}{0.05}
    \setlength{\aboverulesep}{0pt}
    \setlength{\belowrulesep}{0pt}
    \setlength{\abovetopsep}{0pt}
    \setlength{\belowbottomsep}{0pt}
    \setlength{\cmidrulesep}{0pt}
    \setlength{\tabcolsep}{3pt}         
    \setlength{\tabcolsep}{5pt}         
    \caption{Layer size distribution of Docker Hub Top 100 images. Each percentage represents the proportion of layers that are smaller than the specified threshold.}
    \label{table:block_size_dist}
    \begin{tabular}{cr|cr}
    \toprule
    { Thres. (B)} & {\% $<$ Thres.} & {Thres. (B)} & {\% $<$ Thres.}\\
    \midrule
    128 & 1.64 & 1 Ki & 29.27 \\
    8 Ki & 41.45 & 512 Ki & 47.78 \\
    4 Mi & 57.38 & 32 Mi & 76.81 \\
    256 Mi & 97.19 & 605.73 Mi & 100.00 \\
    \bottomrule
    \end{tabular}
\end{table}

The Downloading Engine is structured around a \textit{Request Dispatcher} and a \textit{P2P Downloader}.

\paragraph{Request Dispatcher}
\textcolor{black}{Upon receiving a layer digest from the API interface, the dispatcher first checks with the local Cache Manager (Sec. \ref{s434}) and serves the content if locally present. Otherwise, a three-stage decision process is executed to retrieve the content from external sources. (i) It first queries the other nodes in the same LAN and fetches from the neighbor nodes. (ii) If not local, the dispatcher consults the manifest cache to obtain the layer's size $L_i$ to determine if a P2P process shall be started. If $L_i$ is below a configurable threshold $\theta$ (e.g., 1 MiB), the request is routed directly to the upstream registry. This is justified by the data in Table~\ref{table:block_size_dist}, which shows that a policy of direct fetch for small layers can handle a large fraction of requests with minimal latency. (iii) For layers larger than $\theta$, the dispatcher initiates a P2P discovery process. However, this process is bounded by a dynamic timeout $\tau$, defined as the 95th percentile of recent round-trip times (RTTs) over a 10-second sliding window. If no suitable peer is found within $\tau$, the system falls back to the upstream registry to avoid indefinite user wait times. This logic can be viewed as a state machine (Fig. \ref{partial-p2p-state-machine}), but is now implemented as a streamlined, asynchronous workflow to reduce overhead.}

\begin{figure}[t]
    \includegraphics[width=3.45in]{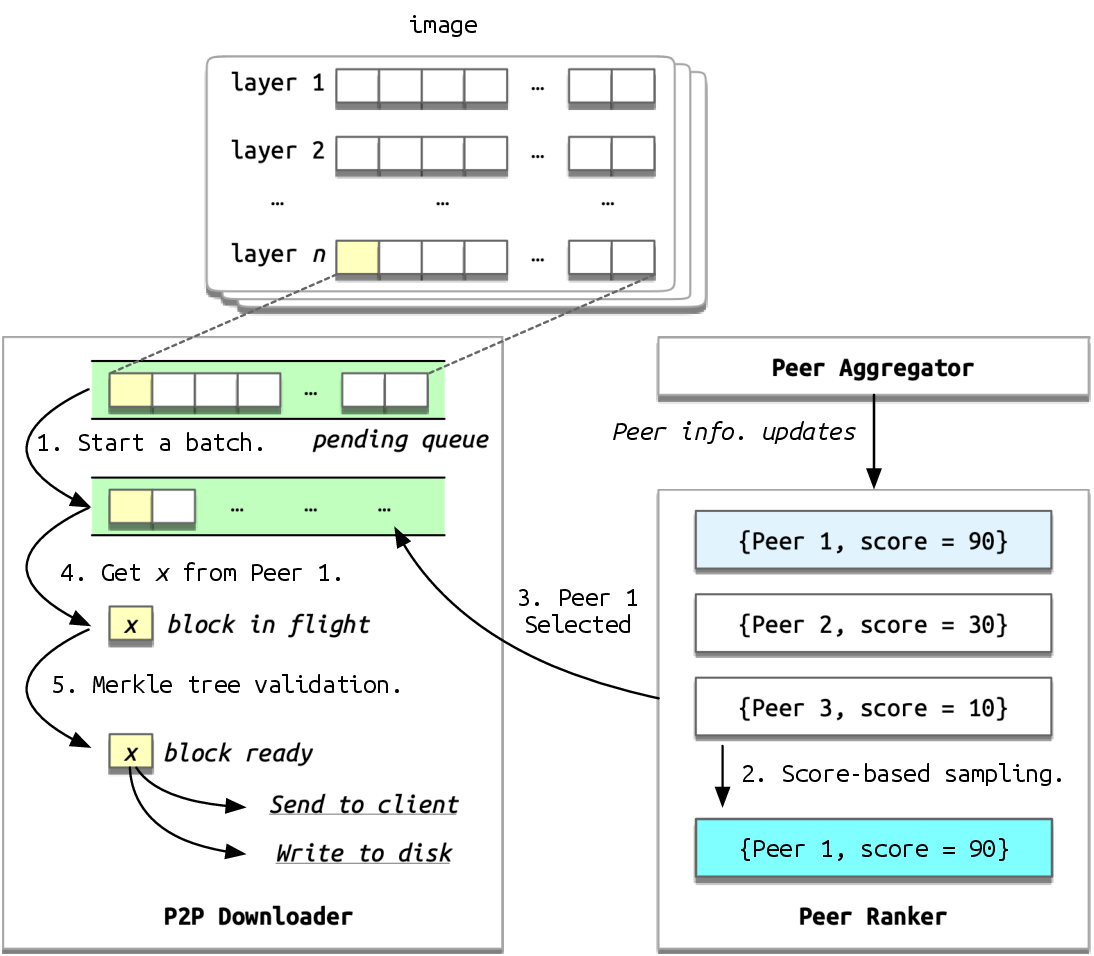}
    \caption{Workflow of P2P downloading in \textsc{PeerSync}.}
    \label{fig_downloader}
\end{figure}

\paragraph{Popularity- and Network-Aware P2P Downloader}
\textsc{PeerSync} segments each image layer into fixed-size \textit{blocks} to enable concurrent downloads from multiple peers. Upon receiving a download request from the Request Dispatcher, the P2P Downloader initiates a five-stage workflow (Fig.~\ref{fig_downloader}): (i) select a batch of blocks for the current cycle; (ii) sample candidate peers based on dynamically updated scores maintained by the Peer Aggregator; (iii) assign each block to the highest-scoring peer; (iv) issue download requests; and (v) validate received blocks via cryptographic hash. During this phase, \textsc{PeerSync} re-queues failed blocks or caching and delivers valid ones.

\begin{table}[h!]
    \centering
    \scriptsize
    \renewcommand{\arraystretch}{0.05}
    \setlength{\aboverulesep}{0pt}
    \setlength{\belowrulesep}{0pt}
    \setlength{\abovetopsep}{0pt}
    \setlength{\belowbottomsep}{0pt}
    \setlength{\cmidrulesep}{0pt}
    \setlength{\tabcolsep}{3pt}         
    \setlength{\tabcolsep}{3pt}         
    \caption{Download times for an 8194.5 MiB image in a 10 Gbps LAN environment with different block sizes.}
    \label{table:block_size}
    \begin{tabular}{ccc}
    \toprule 
    {Block Size (MiB)} & {\#Blocks} & {Download Time (s)}\\
    \midrule 
    256 & 33 & 58\\
    128 & 65 & 59\\
    32 & 257 & 64\\
    16 & 513 & 63\\
    8 & 1025 & 87\\
    \bottomrule 
    \end{tabular}
\end{table}

\textit{Determining the block size.} Block size critically balances parallelism and overhead. Excessively large blocks limit concurrency; overly small ones inflate Merkle tree depth and hash computation costs. Empirical evaluation using an 8194.5 MiB image in a 10 Gbps LAN (Table~\ref{table:block_size}) shows optimal performance at 16-128 MiB. Accordingly, \textsc{PeerSync} sets block size $L_b$ adaptively based on image size $L_i$:
\begin{tightmath}
\begin{equation}
    L_b= \left\{ \begin{array}{ll}
        L_i / 256 & \text{if } L_i \geq 1024 \text{ MiB},\\
        L_i / 64 & \text{if } 256 \text{ MiB} \leq L_i < 1024 \text{ MiB},\\
        L_i / 16 & \text{if } 16 \text{ MiB} \leq L_i < 256 \text{ MiB},\\
        L_i & \text{otherwise.} \\
    \end{array}\right.
    \label{eq_block_size}
\end{equation}
For tiny images whose size $L_i < 16$ MiB, a single block is used to minimize coordination overhead.

\textit{Peer selection rules.} Peer selection combines three criteria into a unified utility score $U(p;t) \in [0,100]$: network position content popularity, and customized logic. 
\begin{enumerate}
    \item \textit{Network position.} Peer selection uses a \textit{network-aware} score that favors peers with higher effective throughput. LAN-local peers receive the maximum score of 100. \textcolor{black}{For remote peers, the score is based on observed download speed, not low-level metrics like packet loss, as congestion control is handled by the OS kernel \cite{cardwell2017bbr,kafi2014congestion,agarwal2023host,arslan2023bolt}; using throughput avoids interfering with kernel mechanisms and better reflects end-to-end performance.}
    Each peer $p$ maintains a sliding window $\mathcal{W}_p$ of past speeds $\{s_p^{t'}\}$. The current speed estimate $s_p^t$ is an exponentially weighted average that prioritizes recent samples:
    \begin{equation}
        s_p^t = \frac{\sum_{s_{p}^{t'} \in \mathcal{W}_p} s_p^{t'} \cdot e^{L - t'} }{\sum_{s_{p}^{t'} \in \mathcal{W}_p} e^{L - t'}},
        \label{eq1}
    \end{equation}
    where $L$ is the current logical time. A global baseline $\bar{s}^t$ is computed similarly over all recent transfers:
    \begin{equation}
        \bar{s}^t = \frac{\sum_{\bar{s}^{t'} \in \mathcal{W}} \bar{s}^{t'} \cdot e^{L - t'} }{\sum_{\bar{s}^{t'} \in \mathcal{W}} e^{L - t'}}.
        \label{eq2}
    \end{equation}
    The raw network advantage is $\text{net}_p^t = s_p^t - \bar{s}^t$, which is then linearly rescaled to $[0,100]$ (with negatives clamped to the minimum known score) to yield the final network score. This relative scoring adapts to heterogeneous and dynamic network conditions.

    \item \textit{Content popularity.} Content popularity discourages reliance on peers holding rare layers. Let $\rho_l \in [0,1]$ denote the fraction of known image instances containing layer $l$:
    \begin{equation}
        \rho_l = \frac{\sum_{p \in \mathcal{P}^t}\sum_{i \in \mathcal{I}_p^t} \epsilon_{l}^{i} }{\sum_{p \in \mathcal{P}^t} \sum_{i \in \mathcal{I}_p^t} 1}, \quad \epsilon_{l}^{i} = 
        \begin{cases}
            1 & \text{if } l \in i,\\
            0 & \text{otherwise}.
        \end{cases}
    \end{equation}
    Peers periodically exchange layer sets via differential updates; unchanged states are acknowledged with sequence numbers to minimize overhead. The popularity score for peer $p$ is then:
    \begin{equation}
        \text{pop}_{p}^t = 100 \times \left( 1 - \frac{\sum_{i \in \mathcal{I}_p^t}\sum_{l \in\mathcal{L}_i} e^{-\lambda \cdot \rho_l}}{\sum_{i \in \mathcal{I}_p^t}\sum_{l \in \mathcal{L}_i} 1} \right),
    \end{equation}
    which down-weights peers storing rarer content. \textcolor{black}{The rationale is to preserve the bandwidth of these critical peers, ensuring they remain available to serve the rare content that only they possess, thus enhancing the overall availability of all layers in the swarm.}

    \item \textit{Extensibility via custom scoring.} \textcolor{black}{To accommodate the diverse and heterogeneous nature of edge deployments, \textsc{PeerSync}'s scoring function is designed with an extensibility hook for domain-specific policies, represented by a custom score $\text{cst}_p^t$. This architectural choice allows administrators to inject logic tailored to their unique operational constraints without modifying the core system. For instance, in a wireless edge environment, an administrator may wish to deprioritize peers with weak signal strength to conserve their battery and bandwidth. This can be achieved by defining the custom score $\text{cst}_p^t$ as a function of the peer's real-time signal strength, thereby penalizing nodes with poor connectivity in the selection process.}
\end{enumerate}
The total utility is a weighted sum:
\begin{equation}
    U(p;t) = \alpha \cdot \text{net}_p^t + \beta \cdot \text{pop}_p^t + \gamma \cdot \text{cst}_p^t,
\end{equation}
with $\alpha + \beta + \gamma = 1$. \textcolor{black}{In the default configuration, $\alpha = \beta = 0.5$, and custom scoring is disabled. The three weights may be adjusted accordingly to achieve a balance between network optimization and resilience. A high $\alpha$ value indicates a strong preference for nearby peers, while $\beta$ strengthens availability. $\gamma$ is configured along with the custom score for more flexibility.} Eventually, peer selection uses a softmax sampling over the utility scores $U(p;t)$, favoring high-utility peers while preserving space for exploration.
\end{tightmath}

\subsubsection{Embedded Autonomous Tracker}\label{s433}
In P2P networks, trackers play a central role in accelerating peer discovery \cite{jia2008designs}. Without an active tracker, nodes must fall back to multi-hop DHT lookups, which significantly increase discovery latency and degrade download performance. However, traditional trackers introduce a single point of failure, vulnerable to crashes, network partitions, or misconfiguration, and require separate deployment and maintenance.

\textcolor{black}{To eliminate this dependency, \textsc{PeerSync} embeds an autonomous tracker module that self-activates when no live trackers are reachable. Specifically, if a node fails to contact any known tracker during bootstrap or periodic health checks, it initiates a leader election within its local network partition using the \textsc{FloodMax} algorithm \cite{10.5555/2821576}.} In this protocol, each candidate broadcasts a stability score (based on metrics including uptime and known neighbors) and propagates the highest score observed so far. The node with the globally maximal score becomes the new tracker. While \textsc{FloodMax} ensures correctness, its naive implementation incurs $O(n^2)$ message complexity in dense networks. To address this, PeerSync applies path-pruning techniques \cite{10.1007/978-981-13-3441-2_2} that suppress redundant message forwarding across already-explored paths and avoid cross-subnet flooding unless necessary. This reduces communication overhead to near-linear complexity while preserving convergence guarantees. 

Since the tracker logic is fully integrated into \textsc{PeerSync}'s runtime, no external configuration or manual intervention is required. This design not only removes operational overhead but also enhances system resilience, which can ensure continuous peer discovery even in transiently partitioned or infrastructure-free edge environments.

\subsubsection{Cache Manager}\label{s434}
Efficient storage management is critical for low-end edge devices, which typically operate under severe resource constraints and rely on flash-based storage without advanced virtualization layers like Ceph \cite{weil2006ceph, makris2022performance}. Over time, \textsc{PeerSync} accumulates cached images, even after their associated applications have terminated, consuming disk space needed for data-intensive tasks such as storing sensor logs or datasets. To preserve system stability while maintaining performance, \textsc{PeerSync} must minimize its storage footprint without compromising image availability. To this end, \textsc{PeerSync} implements a dynamic cache cleaner that monitors image usage via the container runtime and triggers eviction when free space falls below a threshold (e.g., 10\% or a user-defined limit). Eviction decisions are guided by three factors: (i) last access time, (ii) image size, and (iii) content popularity, considering both local presence and global replication.

The underlying algorithm extends the classic Least Recently Used (LRU) policy by incorporating \textit{cache miss cost}, which reflects the non-uniform cost of retrieving an image after eviction \cite{smith1982cache, so1988cache}. Unlike RAM-based caching, in which access latency is uniform, image retrieval in \textsc{PeerSync} varies significantly by network context:
\begin{itemize}
    \item Images with multiple local replicas (within the same LAN) incur near-zero miss cost and are safe to evict.
    \item Images that are the sole local copy are assigned a miss cost inversely proportional to the number of remote replicas.
    \item Sole known copies are assigned the highest retention priority and evicted only under extreme storage pressure.
\end{itemize}

By jointly optimizing recency, space, and retrieval cost, this popularity-aware strategy enables \textsc{PeerSync} to balance disk utilization and performance on resource-constrained edge nodes.

\begin{table*}[t]
\renewcommand{\arraystretch}{0.05}
\setlength{\aboverulesep}{0pt}
\setlength{\belowrulesep}{0pt}
\setlength{\abovetopsep}{0pt}
\setlength{\belowbottomsep}{0pt}
\setlength{\cmidrulesep}{0pt}
\centering
\scriptsize
\caption{The container images chosen for evaluation.}
\label{table:allimage}
\begin{tabular}{cp{3.5cm}crc} 
 \toprule
 {Name} & {Tag} & {Service} & {Compressed Size}& {Description}\\
 \midrule
 redhat/granite-3-1b-a400m-instruct & latest & NLP & 1.47 GB & 1B finetuned IBM Granite 3.0 \cite{granite2024granite} \\
 ai/meta-llama & 3.1-8B-Instruct & NLP & 14.91 GB & Llama 3.1 model by Meta \cite{dubey2024llama}\\
cvisionai/segment-anything & latest & Vision & 5.2 GB & Segment Anything model by Meta \cite{kirillov2023segment}\\
langchain/langchain & latest & NLP & 437.57 MB & LLM application adaptation framework\cite{langchain}\\
pytorch/pytorch & 2.5.1-cuda12.4-cudnn9-runtime & General & 3.11 GB & Deep learning framework\cite{paszke2019pytorch}\\
tensorflow/tensorflow & nightly-gpu & General & 3.61 GB & Deep learning framework\cite{abadi2016tensorflow}\\
 \bottomrule
\end{tabular}
\end{table*}

{\color{black}
\subsection{Implementation Concerns}
\textsc{PeerSync} adopts a message-passing architecture in which each functional component runs as an independent asynchronous task. Components communicate exclusively through well-defined, typed message interfaces, decoupling internal logic from inter-component coordination. A crash or stall in one module (e.g., due to a malformed image manifest or network timeout) does not propagate to others. The supervisor process can restart the failed component within milliseconds, ensuring continuous operation without full-system disruption. In addition, new peer discovery protocols, storage backends, or scoring heuristics can be integrated by implementing standardized interfaces without modifying core logics. This facilitates rapid adaptation to upcoming container formats or edge-specific constraints. 

Peer discovery leverages multiple orthogonal mechanisms, centralized trackers, Kademlia-based DHT, and LAN-local multicast, all operating concurrently. The Downloading Engine monitors availability signals from these sources and initiates block fetching as soon as any peer in the P2P swarm reports possession of the requested layer, mimicking BitTorrent's opportunistic pull model. If no P2P sources are reachable within a configurable timeout (or if the content is globally unique), \textsc{PeerSync} transparently falls back to direct HTTP(S) retrieval from the upstream registry. This hybrid strategy guarantees liveness while maximizing bandwidth utilization whenever collaborative sources exist.
}

{\color{black}
\section{Adaptive Peer Selection and System-Level Guarantees}\label{s41}

Peer selection in \textsc{PeerSync} operates in a non-stationary, partially observable environment where performance and content distribution vary constantly. Rather than pursuing unattainable optimality guarantees against an ill-defined oracle, we analyze the mechanism through three analytically tractable lenses: (i) convergence of utility estimation under exponential smoothing, (ii) probabilistic protection of rare-content peers, and (iii) approximation quality of aggregate throughput maximization. Together, these properties explain why the heuristic consistently achieves high speedup while preserving swarm health.

\subsection{Convergence of Network Utility Estimation}
\begin{tightmath}

The network score $\text{net}_p^t$ is derived from an exponentially weighted moving average (EWMA) of observed download speeds (see \eqref{eq1}), where $\tau_s > 0$ is the smoothing time constant (implicit in your original notation via $L - t'$). Let $B_p(t)$ denote the true instantaneous throughput achievable from peer $p$ at time $t$, and assume $|s_p^{t'} - B_p(t')| \leq \sigma$ for all $t'$ (bounded measurement noise). Then, under mild Lipschitz continuity of $B_p(\cdot)$ (i.e., $|B_p(t) - B_p(t-1)| \leq \delta$), the estimation error satisfies
\begin{equation}
    |s_p^t - B_p(t)| \leq \sigma + \delta \cdot \tau_s.
\end{equation}
Thus, by choosing $\tau_s$ appropriately (e.g., $\tau_s = 2$ rounds in our implementation), the EWMA estimator tracks the true throughput within a bounded error envelope. Since $\text{net}_p^t$ is a monotonic rescaling of $s_p^t - \bar{s}^t$, it inherits this stability, ensuring that transient anomalies do not dominate peer ranking.

\subsection{Probabilistic Protection of Rare-Content Peers}
Let $\mathcal{R}_t = \{ l : \rho_l < \rho_{\min} \}$ denote the set of rare layers at time $t$, and define the rarity exposure of peer $p$ as
\begin{equation}
    \eta_p^t = \frac{1}{|\mathcal{L}_p^t|} \sum_{l \in \mathcal{L}_p^t} \mathbb{I}[l \in \mathcal{R}_t],
\end{equation}
where $\mathcal{L}_p^t$ is the set of layers held by $p$ and $\mathbb{I}[\cdot]$ is the indicator function. The popularity score can be upper-bounded as
\begin{equation}
    \text{pop}_p^t \leq 100 \cdot (1 - e^{-\lambda \rho_{\min}}) \cdot (1 - \eta_p^t) + 100 \cdot (1 - e^{-\lambda}) \cdot \eta_p^t.
\end{equation}
Consequently, if $\eta_p^t \geq \eta_0$ (i.e., peer $p$ holds many rare layers), then $\text{pop}_p^t \leq \bar{u}_{\text{rare}} < 100$. Under softmax selection with temperature $\tau$, the probability of selecting such a peer is upper-bounded by
\begin{align}
    \Pr\{p_t = p\} &\leq \frac{\exp\big((\alpha \cdot 100 + \beta \cdot \bar{u}_{\text{rare}})/\tau\big)}{\exp\big((\alpha + \beta) \cdot 100 / \tau\big)} \nonumber\\ 
    &= \exp\left( -\frac{\beta (100 - \bar{u}_{\text{rare}})}{\tau} \right).
\end{align}
This exponential suppression ensures that peers critical for rare content are selected infrequently, thereby conserving their bandwidth. In steady state, this mechanism bounds the expected depletion rate of rare layers, enhancing long-term availability.

\subsection{Throughput Approximation via Utility Maximization}
Let $\mathcal{S}_t \subseteq \mathcal{P}$ be the set of $k$ peers selected for concurrent download at time $t$. Assume block-level parallelism and negligible coordination overhead, so the total bandwidth is
\begin{equation}
    B_{\text{total}}(t) = B_0 + \sum_{p \in \mathcal{S}_t} B_p(t).
\end{equation}
Since $\text{net}_p^t$ is a linear transformation of $s_p^t$, and $s_p^t$ approximates $B_p(t)$ within error $\epsilon = \sigma + \delta \tau_s$, we have
\begin{equation}
    \left| \sum_{p \in \mathcal{S}_t} \text{net}_p^t - \sum_{p \in \mathcal{S}_t} B_p(t) \right| \leq k \cdot C \cdot \epsilon,
\end{equation}
for some constant $C$ from the rescaling. Because $\text{pop}_p^t \in [0,100]$ is independent of instantaneous bandwidth, the total utility sum satisfies
\begin{equation}
    \sum_{p \in \mathcal{S}_t} U(p;t) = \alpha \sum_{p \in \mathcal{S}_t} \text{net}_p^t + \beta \sum_{p \in \mathcal{S}_t} \text{pop}_p^t + \gamma \sum_{p \in \mathcal{S}_t} \text{cst}_p^t.
\end{equation}
Thus, maximizing $\sum_{p \in \mathcal{S}_t} U(p;t)$ approximately maximizes $\sum_{p \in \mathcal{S}_t} B_p(t)$ up to an additive error of $O(k \epsilon)$. In practice, since $\alpha = \beta = 0.5$, the scheduler jointly optimizes for speed and resilience, achieving a Pareto-efficient trade-off.
\end{tightmath}
}


\section{Evaluation}\label{s5}
We conduct extensive experiments in two environments: a Docker Compose-based emulation on a high-performance x86\_64 host (Intel Xeon Silver 4214) and physical edge devices (Raspberry Pi 4 Model B). The emulation enables the simulation of large-scale environments, while the physical setup validates the practical feasibility of \textsc{PeerSync}. We select a range of popular AI/ML and commonly used applications of varying sizes, from small base images to large language model (LLM) containers (Table~\ref{table:allimage}). These images represent diverse use cases, allowing us to evaluate system performance under varying workloads.

We evaluated \textsc{PeerSync} against two popular container image distribution systems: \textit{Dragonfly}~\cite{d7o} and \textit{Kraken}~\cite{kraken}, as well as the plain HTTP-based pull method, referred to as the \textit{Baseline}.
\begin{itemize}
    \item Dragonfly, hosted by the CNCF as an Incubating Level Project, leverages P2P technology to accelerate content distribution. It is designed to enhance efficiency in distributed environments by reducing reliance on centralized registries.
    \item Kraken, an open-source project developed by Uber and deployed in production since early 2018, also employs P2P technology for Docker image management, replication, and distribution. Kraken is particularly tailored for hybrid cloud environments, offering decentralized capabilities to optimize large-scale image deployments.
\end{itemize}
Although Starlight~\cite{278314} also focuses on improving container image delivery, it relies on fundamentally different mechanisms, such as runtime trace collection and dependence on centralized registries. These differences make direct comparisons with \textsc{PeerSync} less relevant (see Sec.~\ref{s3}). Consequently, we did not include Starlight in our comparisons.


\begin{figure*}[t]
    \begin{minipage}{0.5\textwidth}
        \centering
        \includegraphics[width=\textwidth]{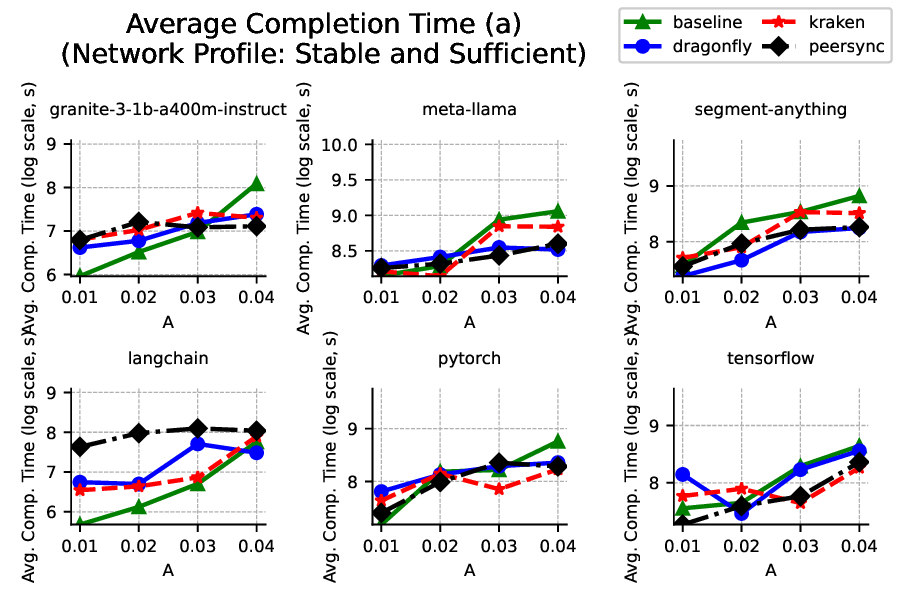}
    \end{minipage}%
    \hfill
    \begin{minipage}{0.5\textwidth}
        \centering
        \includegraphics[width=\textwidth]{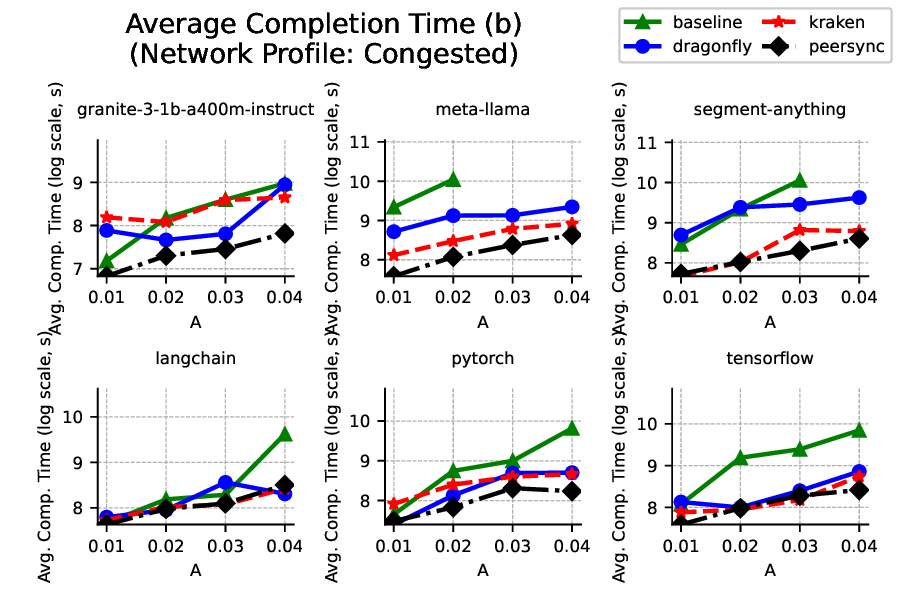}
    \end{minipage}
    
    \begin{minipage}{0.5\textwidth}
        \centering
        \includegraphics[width=\textwidth]{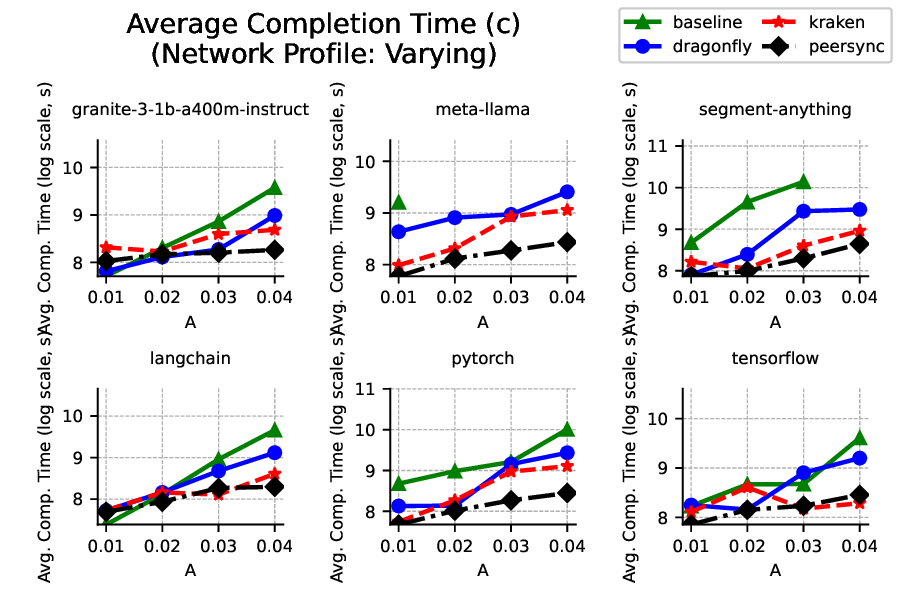}
    \end{minipage}%
    \hfill
    \begin{minipage}{0.5\textwidth}
        \centering
        \includegraphics[width=\textwidth]{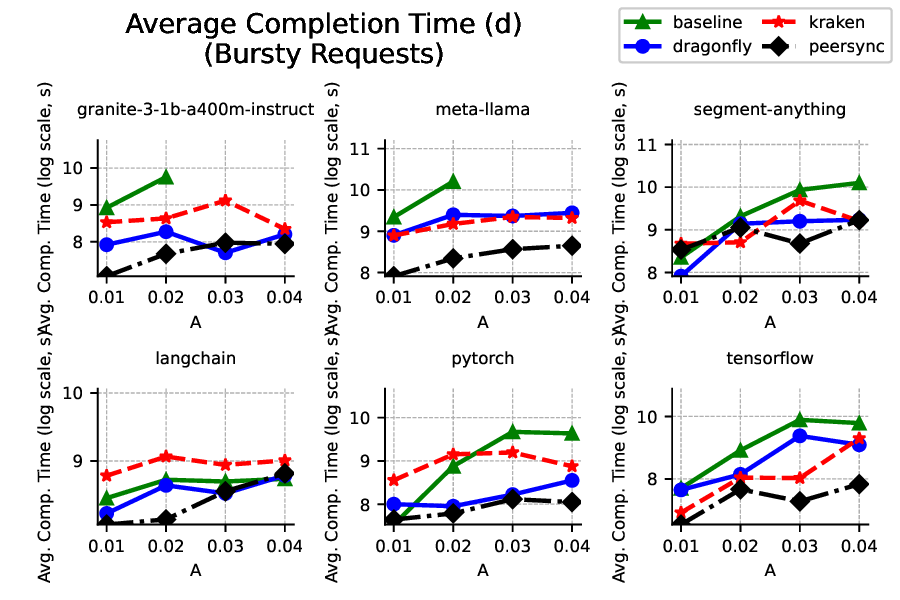}
    \end{minipage}
    
    \caption{Average AI/ML image distribution time under various network conditions (log scale). The $x$-axis is the parameter $A$. The missing points mean that the download operation could not be completed during the limited time frame.}
    \label{fig_ai}
\end{figure*}




\subsection{Docker Compose-based Emulation}\label{s5.1}
\subsubsection{Experimental Setup}
We deployed a Docker Compose-based emulation environment comprising 10 Linux bridge networks (\texttt{net\_worker\{$n$\}}, $n = 1$ to $10$) to model a distributed system of interconnected edge sites. Each network hosted one router responsible for inter-network communication and seven worker nodes issuing container image pull requests. Routers were configured using \texttt{tc(8)}~\cite{tc} to impose realistic WAN characteristics, including bandwidth limits (20 Mbps to 500 Mbps), variable latency, and packet loss, mimicking ISP and transit links. In contrast, intra-network communication within each LAN was provisioned with uncapped bandwidth and zero packet loss to reflect typical local cluster conditions. \textcolor{black}{This multi-LAN environment models the Internet, where LANs are edge sites and routers are ISPs. In each edge site, traffic is routed internally without flowing through external ISPs.}

To emulate heterogeneous user demand, we modeled image request arrival times using an inverse Poisson process:
\[
t_i \sim \text{Poisson}^{-1}\left( \text{random} \left(0.001, A \cdot e^{\frac{B}{\text{s}_i}} \right)\right),
\]
where $\text{s}_i$ denotes the size of image $i$, and $A$, $B$ are tunable parameters controlling request intensity. Larger values of $A$ and $B$ correspond to higher request frequencies, simulating peak-load scenarios. Background traffic generated via \texttt{iPerf}~\cite{iperf} introduced additional congestion on shared links, better reflecting real-world multi-tenant network usage. All centralized services, including container registries, Dragonfly metadata databases, and Kraken trackers, were co-located in \texttt{net\_worker1} to emulate a cloud-hosted control plane.

\textcolor{black}{For reproducibility of our evaluation, Table \ref{table:emu_params} gives full parameters used in our emulation. These parameters include values for network bandwidth and packet loss, node CPU and RAM allocation, and how we configure \textsc{PeerSync} to balance between network and popularity-aware scoring. For the Varying Network experiment group, the bandwidth and packet loss values are randomly modified within the range specified.}

\begin{table}[h!]
    \scriptsize
    \centering
    \renewcommand{\arraystretch}{0.05}
    \setlength{\aboverulesep}{0pt}
    \setlength{\belowrulesep}{0pt}
    \setlength{\abovetopsep}{0pt}
    \setlength{\belowbottomsep}{0pt}
    \setlength{\cmidrulesep}{0pt}
    \setlength{\tabcolsep}{3pt}         
    \caption{\color{black}Key parameters used in the emulation.}
    \label{table:emu_params}
    \begin{tabular}{lc}
    \toprule
    {Parameter} & {Value}\\
    \midrule
    Number of LANs & 10\\
    Number of Nodes per LAN & 7\\
    Emulation Timespan & 1800 seconds\\
    Bandwidth (Internal) & Uncapped\\
    Packet Loss (Internal) & 0\%\\
    Bandwidth (Stable Net.) & Uncapped\\
    Packet Loss (Stable Net.) & 0\%\\
    Bandwidth (Congested Net.) & 100 to 500Mbps\\
    Packet Loss (Congested Net.) & 0 to 10\%\\
    Bandwidth (Varying Net., Min) & 20 to 150Mbps\\
    Packet Loss (Varying Net., Min) & 0 to 10\%\\
    Bandwidth (Varying Net., Max) & 100 to 500Mbps\\
    Packet Loss (Varying Net., Max) & 30 to 50\%\\
    Peer Churn Ratio (Varying Net.) & 10\%\\
    Environmental Variation Cycle (Varying Net.) & 60 seconds\\
    Node CPU Allocation & 2 to 4 shared cores\\
    Node RAM Allocation & 2 to 4 GiB\\
    \textsc{PeerSync} Scoring Weight $\alpha$ & 0.5\\
    \textsc{PeerSync} Scoring Weight $\beta$ & 0.5\\
    \bottomrule
    \end{tabular}
\end{table}



\subsubsection{Image Distribution Time}  
Figure~\ref{fig_ai} presents the average container image distribution times under different network conditions as a function of request frequency (controlled by parameter $A$). Due to the wide dynamic range, especially under high load, a base-2 logarithmic scale is used on the $y$-axis for clarity.

\begin{table}[h!]
    \centering
    \renewcommand{\arraystretch}{0.05}
    \setlength{\aboverulesep}{0pt}
    \setlength{\belowrulesep}{0pt}
    \setlength{\abovetopsep}{0pt}
    \setlength{\belowbottomsep}{0pt}
    \setlength{\cmidrulesep}{0pt}
    \scriptsize
    \caption{\textcolor{black}{Normalized average completion time (Baseline = 100\%, lower is better) under different network profiles.}}
    \label{table:speedup}
    \begin{tabular}{r|rrrr}
    \toprule
    {Profile} & {Baseline} & {Dragonfly \cite{d7o}} & {Kraken \cite{kraken}} & \textsc{PeerSync}\\
    \midrule
    Stable & 100\% & 75.01\% & 80.13\% & \textbf{77.85\%}\\
    Congested & 100\% & 64.27\% & 61.89\% & \textbf{45.89\%}\\
    Varying & 100\% & 65.63\% & 46.96\% & \textbf{36.71\%}\\
    \bottomrule
    \end{tabular}
\end{table}

\paragraph{Stable network conditions (Fig. \ref{fig_ai} (a))}  
In environments with sufficient and stable network resources, all P2P-based methods, including \textsc{PeerSync}, exhibit similar performance trends as request frequency increases (see Table \ref{table:speedup}). While distribution latency increases slightly with higher request loads, P2P methods consistently outperform the Baseline HTTP approach. This aligns with the known benefits of P2P models, such as efficient data sharing and reduced reliance on centralized registries that have been extensively validated in data center contexts. The trend is particularly pronounced for large LLM images: the average distribution time roughly quadruples as request frequency increases, highlighting the critical role of P2P technologies in facilitating the deployment of modern AI/ML models with billions of parameters.

\paragraph{Congested network conditions (Fig.~\ref{fig_ai} (b))}  
In environments with bandwidth limitations, packet loss, and high latency, the Baseline method suffers severe performance degradation due to the long-tail effects of centralized pulling. Remote data transfers are particularly affected, often exceeding the default time limit of 1200 seconds, as indicated by the missing points in the green Baseline lines. In contrast, \textsc{PeerSync} and other P2P-based methods maintain better performance by distributing the load across multiple peers, effectively mitigating bandwidth constraints. Notably, Dragonfly shows a sharp performance drop for larger images due to its reliance on centralized components such as schedulers and managers. As communication between edge nodes and these components becomes unreliable, Dragonfly’s ability to coordinate cloud-edge interactions weakens, resulting in bottlenecks during image serving. \textsc{PeerSync}, by contrast, leverages its decentralized architecture to achieve consistent performance, even under congestion.

\paragraph{Variable network conditions (Fig.~\ref{fig_ai} (c))}  
This scenario simulates real-world edge environments where latency, packet loss, and bandwidth fluctuate significantly, and participating nodes frequently join and leave the network. \textcolor{black}{The corresponding parameters are given in Table \ref{table:emu_params}. In each variation cycle, every emulated router adopts a new random configuration within the predefined range, 10\% of the currently active peers leave the swarm, and a subset of previously disconnected peers rejoin the swarm.} Such instability disrupts inter-peer connections and hinders image fetching, even for smaller images like \texttt{langchain/langchain}. Compared to other P2P-based solutions, \textsc{PeerSync} demonstrates superior resilience under these conditions. By prioritizing local peer synchronization and minimizing dependence on centralized components, \textsc{PeerSync} reduces reliance on specific, potentially unstable links and dynamically explores all available links to maximize performance. This strategy not only improves distribution efficiency but also enhances fault tolerance, which is critical in environments with unpredictable network behavior.

\textsc{PeerSync}'s ability to optimize local storage and peer communication makes it particularly robust in dynamic edge environments. As shown in Table~\ref{table:speedup}, \textsc{PeerSync} outperforms the Baseline, Dragonfly, and Kraken under unstable, high-frequency scenarios by average factors of 2.72, 1.79, and 1.28, respectively.\footnote{The timeout for each request was 1200 seconds. In some instances, the Baseline could not finish within the limit. We assigned 1200 seconds to the missing data points to avoid missing points. Consequently, the actual speedup relative to the Baseline is higher than the calculated value.} The speedup is even more pronounced for larger container images, reinforcing \textsc{PeerSync}'s effectiveness in addressing the challenges of large-scale image distribution.

\subsubsection{Uplink Occupancy}
Tables \ref{table:cross_stable}, \ref{table:cross_congested}, and \ref{table:cross_var} illustrate the cross-network traffic for each image distribution method across three distinct network profiles: stable, congested, and unstable conditions. Cross-network traffic represents data that traverses routers and encounters bandwidth limits, latency, and packet loss, effectively simulating the challenges of data traversing upstream ISP transit networks and the Internet. Unlike local network bandwidth, which is often abundant, cross-network bandwidth is typically constrained by external factors such as physical link limitations and tariff plans. Thus, minimizing cross-network traffic is critical, particularly in edge computing, where the goal is to prioritize intra-LAN resources as much as possible.

\begin{table}[h!]
    \centering
    \renewcommand{\arraystretch}{0.05}
    \setlength{\aboverulesep}{0pt}
    \setlength{\belowrulesep}{0pt}
    \setlength{\abovetopsep}{0pt}
    \setlength{\belowbottomsep}{0pt}
    \setlength{\cmidrulesep}{0pt}
    \setlength{\tabcolsep}{3pt}         
    \scriptsize
    \caption{Aggregate inter-LAN traffic under Stable profile.}
    \label{table:cross_stable}
    \begin{tabular}{crr} 
     \toprule
     {Solution} & {Maximum (Gbps)} & {Average (Gbps)} \\
     \midrule
     \textsc{PeerSync} & \textbf{5.07} & \textbf{0.82}\\
     Kraken \cite{kraken} & 6.61 & 0.93\\
     Dragonfly \cite{d7o} & 8.80 & 1.37\\
     Baseline & 11.91 & 6.18\\
     \bottomrule
    \end{tabular}
\end{table}

\begin{table*}[t]
    \centering
    \renewcommand{\arraystretch}{0.05}
    \setlength{\aboverulesep}{0pt}
    \setlength{\belowrulesep}{0pt}
    \setlength{\abovetopsep}{0pt}
    \setlength{\belowbottomsep}{0pt}
    \setlength{\cmidrulesep}{0pt}
    \scriptsize
    \caption{Number of devices in a LAN and the average distribution time (over 100 retrieval requests) served by LAN cache.}
    \label{table:cache1}
    \begin{tabular}{c|rrrrrrrrrr} 
     \toprule
     {Number of edge devices within a LAN} & 1 & 2 & 3 & 4 & 5 & 6 & 7 & 8 & 9 & 10\\
     \midrule
     {Average image distribution time (s)} & 9.11 & 8.33 & 9.69 & 11.34 & 8.87 & 8.06 & 5.08 & 2.86 & 2.89 & 2.19\\
     \bottomrule
    \end{tabular}
\end{table*}

\begin{itemize}
    \item \textit{Sufficient and stable network conditions.} As shown in Table \ref{table:cross_stable}, \textsc{PeerSync} consistently minimizes cross-network bandwidth usage under stable conditions. By prioritizing local peer data, \textsc{PeerSync} achieves an average bandwidth consumption of only 0.82 Gbps, significantly lower than Kraken (0.93 Gbps), Dragonfly (1.37 Gbps), and the Baseline (6.18 Gbps). While Dragonfly and Kraken leverage P2P techniques, their reliance on centralized components for coordination and scheduling results in higher cross-network traffic. The Baseline method incurs the highest bandwidth usage due to its centralized pulling mechanism, which does not take advantage of local data sharing.
    \item \textit{Congested but stable network conditions.} In congested but stable network conditions (Table \ref{table:cross_congested}), \textsc{PeerSync} continues to outperform other methods by maintaining an average cross-network traffic of just 0.91 Gbps. The Baseline approach shows the lowest maximum bandwidth (10.36 Gbps), which is a result of its slower and less efficient data retrieval. However, it saturates network capacity with an average bandwidth usage of 9.81 Gbps, underscoring its inefficiency in managing congested networks. Kraken and Dragonfly perform better than the Baseline but still exhibit performance degradation due to their centralized dependencies. Dragonfly, in particular, suffers from its reliance on schedulers and managers, resulting in an average traffic of 4.69 Gbps, over five times higher than \textsc{PeerSync}.
    \item \textit{Congested and unstable network conditions.} Table \ref{table:cross_var} highlights the performance of each method under congested and unstable network conditions, characterized by fluctuating packet loss, latency, and bandwidth. \textsc{PeerSync} demonstrates remarkable robustness, maintaining the lowest average cross-network traffic at 0.76 Gbps. Kraken performs slightly worse, with an average traffic of 0.89 Gbps, but still manages to leverage its decentralized architecture to reduce reliance on external networks. Dragonfly, on the other hand, suffers significant performance degradation due to its dependence on centralized coordination, resulting in an average traffic of 4.01 Gbps. The Baseline method experiences the lowest maximum bandwidth usage (8.87 Gbps) but continues to exhibit high average traffic (6.34 Gbps), reflecting its inability to adapt to varying network conditions. This highlights the inefficiency of centralized pulling in handling unstable and dynamic environments.
\end{itemize}


\begin{table}[h!]
    \centering
    \renewcommand{\arraystretch}{0.05}
    \setlength{\aboverulesep}{0pt}
    \setlength{\belowrulesep}{0pt}
    \setlength{\abovetopsep}{0pt}
    \setlength{\belowbottomsep}{0pt}
    \setlength{\cmidrulesep}{0pt}
    \setlength{\tabcolsep}{3pt}         
    \scriptsize
    \caption{Aggregate inter-LAN traffic under Congested profile.}
    \label{table:cross_congested}
    \begin{tabular}{crr} 
     \toprule
     {Solution} & {Maximum (Gbps)} & {Average (Gbps)} \\
     \midrule
     \textsc{PeerSync} & 11.50 & \textbf{0.91}\\
     Kraken \cite{kraken} & 13.76 & 1.18\\
     Dragonfly \cite{d7o} & 11.94 & 4.69\\
     Baseline & \textbf{10.36} & 9.81\\
     \bottomrule
    \end{tabular}
\end{table}

\begin{table}[h!]
    \centering
    \scriptsize
    \renewcommand{\arraystretch}{0.05}
    \setlength{\aboverulesep}{0pt}
    \setlength{\belowrulesep}{0pt}
    \setlength{\abovetopsep}{0pt}
    \setlength{\belowbottomsep}{0pt}
    \setlength{\cmidrulesep}{0pt}
    \setlength{\tabcolsep}{3pt}         
    \caption{Aggregate inter-LAN traffic under Varying profile.}
    \label{table:cross_var}
    \begin{tabular}{c|rr} 
     \toprule
     {Solution} & {Maximum (Gbps)} & {Average (Gbps)} \\
     \midrule
     \textsc{PeerSync} & 9.15 & \textbf{0.76}\\
     Kraken \cite{kraken} & 10.99 & 0.89\\
     Dragonfly \cite{d7o} & 10.13 & 4.01\\
     Baseline & \textbf{8.87} & 6.34\\
     \bottomrule
    \end{tabular}
\end{table}


The results across all network profiles clearly demonstrate \textsc{PeerSync}'s superior ability to minimize cross-network bandwidth consumption. This efficiency is achieved by leveraging local resources and reducing reliance on external network traffic, which is particularly critical in edge environments where bandwidth is constrained and often unstable.


\begin{table*}[h!]
\centering
\scriptsize
\caption{Number of nodes (1 $\sim$ 10) and comparison of the sum of cache space occupied by Cache Cleaner and the LRU policy.}
\label{table:cache_lru1}
\begin{tabular}{c|rrrrrrrrrr} 
 \toprule
 {Number of edge devices within a LAN} & 1 & 2 & 3 & 4 & 5 & 6 & 7 & 8 & 9 & 10\\
 \midrule
 {Space occupied by Cache Cleaner (MiB)} & 196 & 388 & 475 & 589 & 593 & 668 & 905 & 1259 & 1199 & 1420\\
 {Space occupied by LRU (MiB)} & 110 & 206 & 411 & 701 & 837 & 948 & 973 & 1164 & 1056 & 1701\\
 \bottomrule
\end{tabular}
\end{table*}

\begin{figure*}[t]
    \centering
    \includegraphics[width=\textwidth]{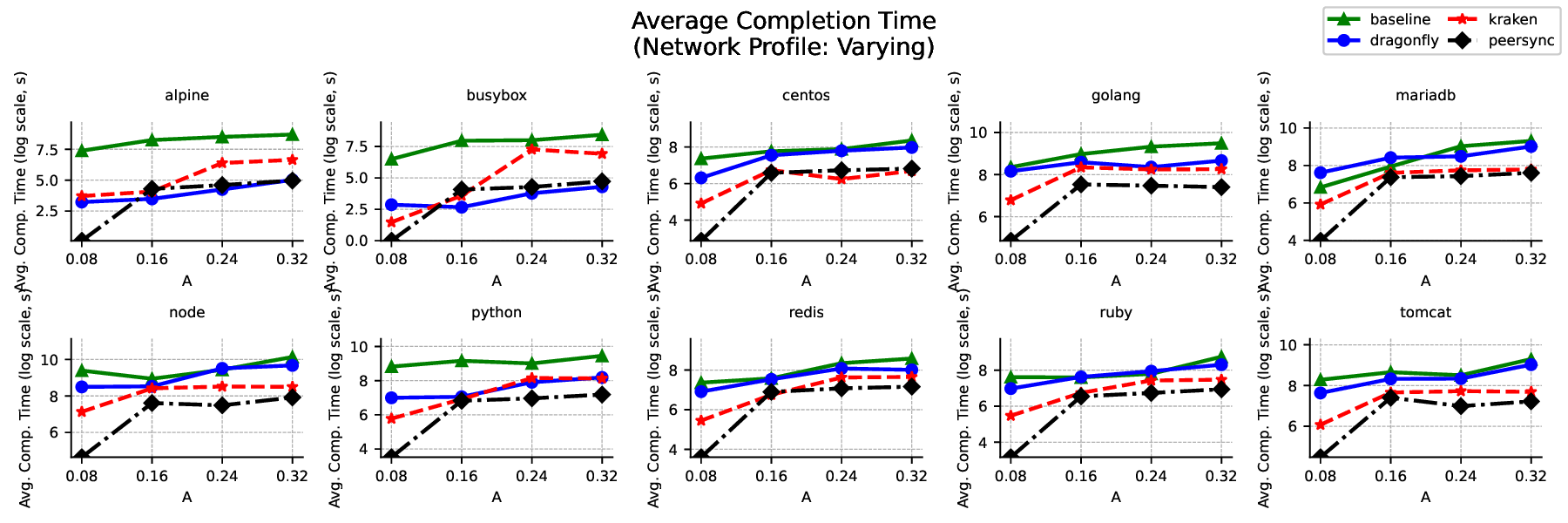}
    \caption{Average image distribution time under Varying profile (log scale). The $x$-axis is the parameter $A$.}
    \label{fig_data_var}
\end{figure*}

\subsubsection{Cache Strategy}
We evaluated the Cache Cleaner algorithm by analyzing its effectiveness in managing requests and accelerating image retrieval within a LAN. Table \ref{table:cache1} shows the relationship between the number of edge devices in a LAN and the average time to download images. Initially, as the number of nodes increases, image distribution times rise slightly due to simultaneous requests to the upstream registry. However, as the LAN becomes more populated, the local cache network is increasingly leveraged, resulting in significantly reduced average retrieval times.

As nodes in the LAN form a collaborative cache, the Cache Cleaner algorithm effectively retains images that are likely to benefit the entire network. This approach optimizes local data sharing and minimizes redundant requests to external networks. The results demonstrate that as the cache network scales, the average download time decreases dramatically, underscoring the efficiency of \textsc{PeerSync}'s collaborative caching mechanism.

To further validate the Cache Cleaner's performance, we compared its total cache usage against the Least Recently Used (LRU) policy in Table \ref{table:cache_lru1}. Cache Cleaner consistently achieves better space utilization in multi-node settings by coordinating cache management across neighboring peers. This approach avoids redundant caching of identical content across multiple nodes, enabling more efficient storage utilization within the LAN. In contrast, LRU operates independently and lacks collaboration, leading to higher total cache consumption. Cache Cleaner’s ability to prioritize fast intra-LAN retrieval not only optimizes storage but also improves overall efficiency.

\subsubsection{Distribution of smaller images} Although our primary focus is on large AI/ML images, smaller, non-AI base images are also widely used in practical setups. To emulate the operation of Docker Hub, which serves popular images in large volumes daily, we selected the 10 most downloaded images from Docker Hub and increased the request frequency by adjusting the parameter $A$. Fig. \ref{fig_data_var} presents the results, demonstrating \textsc{PeerSync}'s ability to maintain its performance advantage over other P2P-based solutions, even under high-frequency request conditions for smaller images. These results highlight \textsc{PeerSync}'s capability to handle the significant volumes typical of Docker Hub operations, proving its versatility across diverse workloads.
\subsubsection{\textcolor{black}{Bursty requests}} \textcolor{black}{In addition to the experiment settings that assume a Poisson arrival process, we evaluate \textsc{PeerSync} under a bursty request-arrival scenario, where the request arrival rate temporarily increases. Additionally, we mimic diurnal request patterns by applying bursts of different magnitudes. More specifically, this setting is based on the Congested Network AI/ML image distribution experiment, except that the Poisson rate parameter is varied periodically. The arrival rate increases to $1.5\times$ for 20\% of the emulation period, and $3.0\times$ for another 10\%, thus periodically applying bursty pressure to the swarm. The evaluation results are given in Fig. \ref{fig_ai} (d). Empirical results show that \textsc{PeerSync} is able to maintain consistent high service quality under such bursty and diurnal request arrival patterns.}

\subsection{Real-World Experiments}\label{s5.2}
\begin{figure}[!ht]
    \centering
    \includegraphics[width=2.75in]{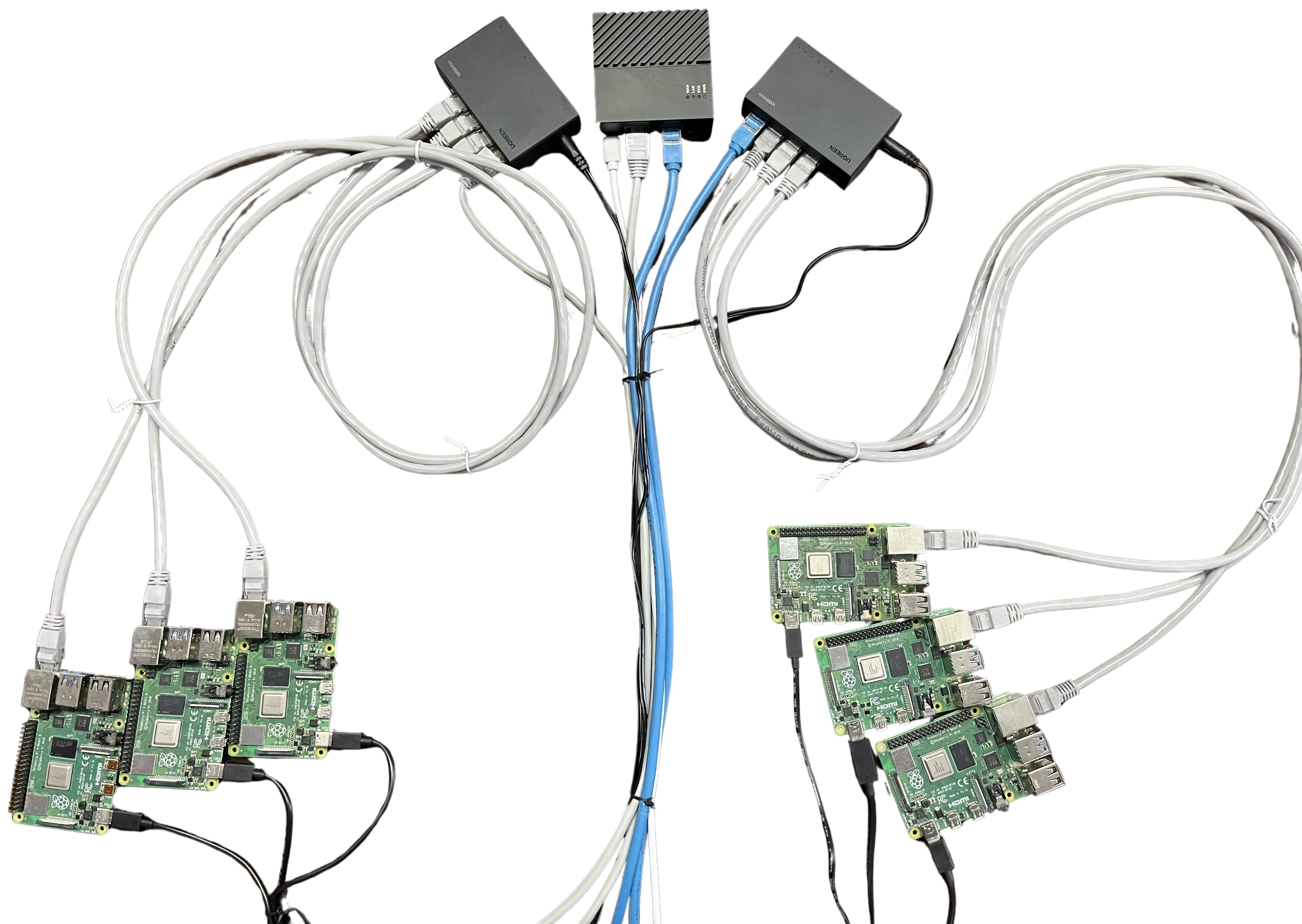}
    \caption{The testbed for real-world experiments.}
    \label{real_world_exp}
\end{figure}

We deployed a physical testbed using Raspberry Pi devices to compare \textsc{PeerSync}'s performance against the Baseline, Dragonfly, and Kraken. As shown in Fig. \ref{real_world_exp}, the setup comprises six Raspberry Pi (RPi) 4 Model B devices connected via two layer 2 switches, each with a 1 Gbps link speed. Each LAN contains three RPis, with the two switches connected to a common router. The router is configured to forward packets between the networks while limiting the inter-LAN bandwidth to 100 Mbps. To simplify deployment, static routes are configured on the router. This setup is similar to the Docker Compose-based emulation described earlier, with two key differences: (i) Physical hardware and real-world networking scenarios are used; (ii) The aarch64 version of \textsc{PeerSync} is deployed instead of the amd64 version to accommodate the ARM-based Raspberry Pi.

Benchmark requests were made to agents running on each RPi to measure image distribution times. Table \ref{table:rpi} presents the 90th and 99th percentiles of distribution times achieved by each system. The results demonstrate that \textsc{PeerSync} significantly outperforms the other systems in real-world conditions.

\begin{itemize}
    \item At the 90th percentile (P90), \textsc{PeerSync} achieves an image distribution time of 190.79 seconds, representing a 39.7\% improvement over Kraken, a 60.9\% improvement over Dragonfly, and a 75.9\% improvement over Baseline.
    \item At the 99th percentile (P99), \textsc{PeerSync} maintains its advantage, with a distribution time of 243.89 seconds, which is 27.8\% faster than Kraken, 53.9\% faster than Dragonfly, and 69.9\% faster than the Baseline.
\end{itemize}

The results show that \textsc{PeerSync} performs better than the other methods significantly in our real-world experiments.

\begin{table}[h!]
    \centering
    \scriptsize
    \renewcommand{\arraystretch}{0.5}  
    \setlength{\tabcolsep}{3pt}         
    \caption{The 90th and 99th percentile of distribution times.}
    \label{table:rpi}
    \begin{tabular}{c|rrrrr} 
     \toprule
     {Solution} & {Baseline} & {Dragonfly} & {Kraken} & \textsc{PeerSync} \\
     \midrule
     P90 (s) & 790.13 & 487.50 & 316.42 & \textbf{190.79} \\
     P99 (s) & 811.23 & 528.49 & 337.75 & \textbf{243.89} \\
     \bottomrule
    \end{tabular}
\end{table}

\section{Related Works}\label{s6}
Numerous solutions have been explored to enhance delivery speeds for container images. In this section, we briefly review key developments in this area, covering both container-specific works and methodologies for other resource types.

Before the era of edge computing, several pioneering works focused on fast resource delivery in data center networks. VDN \cite{peng2012vdn} leverages hierarchical network topology and shared chunks between virtual machine images to achieve a 30-80$\times$ speedup for large virtual machine images under heavy traffic in data center networks. Similarly, VMTorrent \cite{reich2012vmtorrent} improves P2P methods used in live streaming, employing block prioritization, profile-based execution prefetch, on-demand fetch, and decoupling of virtual machine image presentation from the underlying data stream. Experiments show VMTorrent can achieve a 30x speedup over traditional network storage.
Since the rise of Docker, containers have become central to deploying workloads in data centers, and distributing container images has emerged as a key issue. While similar to virtual machine images, containers have unique properties, such as a layered structure and registry-based hosting. DevOps practices demand more automated and integrated workload deployment solutions. FID \cite{kangjin2017fid}, an early P2P-based approach for fast container image distribution, adapts BitTorrent to provide an integrated distribution system. In industry, Dragonfly \cite{d7o} and Kraken \cite{kraken} are prominent P2P container image distribution systems. These methods outperform traditional approaches by utilizing spare bandwidth between clients.

However, these solutions do not explicitly address edge computing environments. While Dragonfly \cite{d7o} and Kraken \cite{kraken} are P2P-based, they rely on multiple centralized components, which can degrade service quality in the event of component failure. Edge environments are less stable than traditional cloud facilities, and deploying these centralized components in the cloud requires manual configuration across different edge locations. Additionally, edge devices typically have weaker network capabilities, so unrestricted use of P2P protocols can lead to uplink congestion. EdgePier \cite{becker2021edgepier}, based on IPFS \cite{benet2014ipfs}, provides a fully decentralized container image distribution system tailored to edge computing. Similarly, Gazzetti et al. \cite{gazzetti2018scalable} proposed a decentralized solution with managers that compute optimal network topologies. \textcolor{black}{For storage-restricted edge devices, learning-based intelligent caching effectively reduces further container spawn latency \cite{10637588, 10437757, torabi2024learning}. While we also employ this incentive in our work, we focus more on providing a full-lifecycle image management solution for systematic optimization of container image management.}
\textcolor{black}{Recent advancements in edge computing offloading, exemplified by systems like FlexSlice \cite{mohajer2024dynamic}, employ RL/TD3 techniques to generate optimal scheduling decisions in fluctuating scenarios. In contrast, \textsc{PeerSync} is designed with heuristic approaches that do not involve online learning. This decision is motivated by the significant operational requirements of learning-based models, including access to large training datasets, specialized GPU/NPU, and stable feedback loops that are frequently unavailable in typical edge deployments. Consequently, \textsc{PeerSync}'s use of a sliding window and RTT-based metrics provides a computationally efficient solution that achieves near-optimal locality, making it better suited for latency-sensitive, one-shot pull operations. }

Beyond P2P, another trend involves distributed storage to form a shared storage layer among participating nodes. CoMICon \cite{nathan2017comicon}, Cider \cite{du2017cider}, and Wharf \cite{zheng2018wharf} use distributed filesystems to accelerate container provisioning. However, these methods are primarily focused on data centers, and implementing distributed storage in edge environments remains challenging \cite{makris2022towards}.

More recently, Starlight \cite{278314} has introduced a novel approach by redesigning the container image architecture. It transmits only the necessary components, significantly reducing transfer times. However, Starlight's reliance on specific container engines limits its applicability, and it continues to use a traditional client-server registry architecture, which restricts throughput under high bandwidth loads. Future work could integrate Starlight with \textsc{PeerSync} to achieve distributed transmission of large data chunks and selective transmission of necessary data, further improving the containerization ecosystem.

\section{Concluding Remarks}\label{s7}
In this paper, we introduced \textsc{PeerSync}, a fully decentralized image distribution system tailored for containerized model inference at the network edge. \textsc{PeerSync} leverages P2P downloading to dynamically adapt to changing network conditions and content popularity, significantly outperforming traditional approaches. Its autonomous tracker eliminates single points of failure, enhancing resilience in unstable and resource-constrained edge environments. The integrated cache cleaner also ensures efficient storage use without compromising performance.

\textcolor{black}{Building on the modular approach, future work could also incorporate intelligent caching as on-demand modules. This would allow for dynamic adaptation to various edge site designs, for example, by implementing proactive caching in predictable environments where rich metrics are available.}
Another promising direction is integrating \textsc{PeerSync} with Starlight~\cite{278314}. Combining \textsc{PeerSync}'s P2P distribution with Starlight’s partial transmission and novel image format could enable a next-generation container ecosystem optimized for edge model inference.
\ifCLASSOPTIONcaptionsoff
  \newpage
\fi

\bibliographystyle{IEEEtran}
\bibliography{ref}

\vspace{-1.1cm}

\begin{IEEEbiography}
  [{\includegraphics[width=1in,height=1.25in,clip,keepaspectratio]{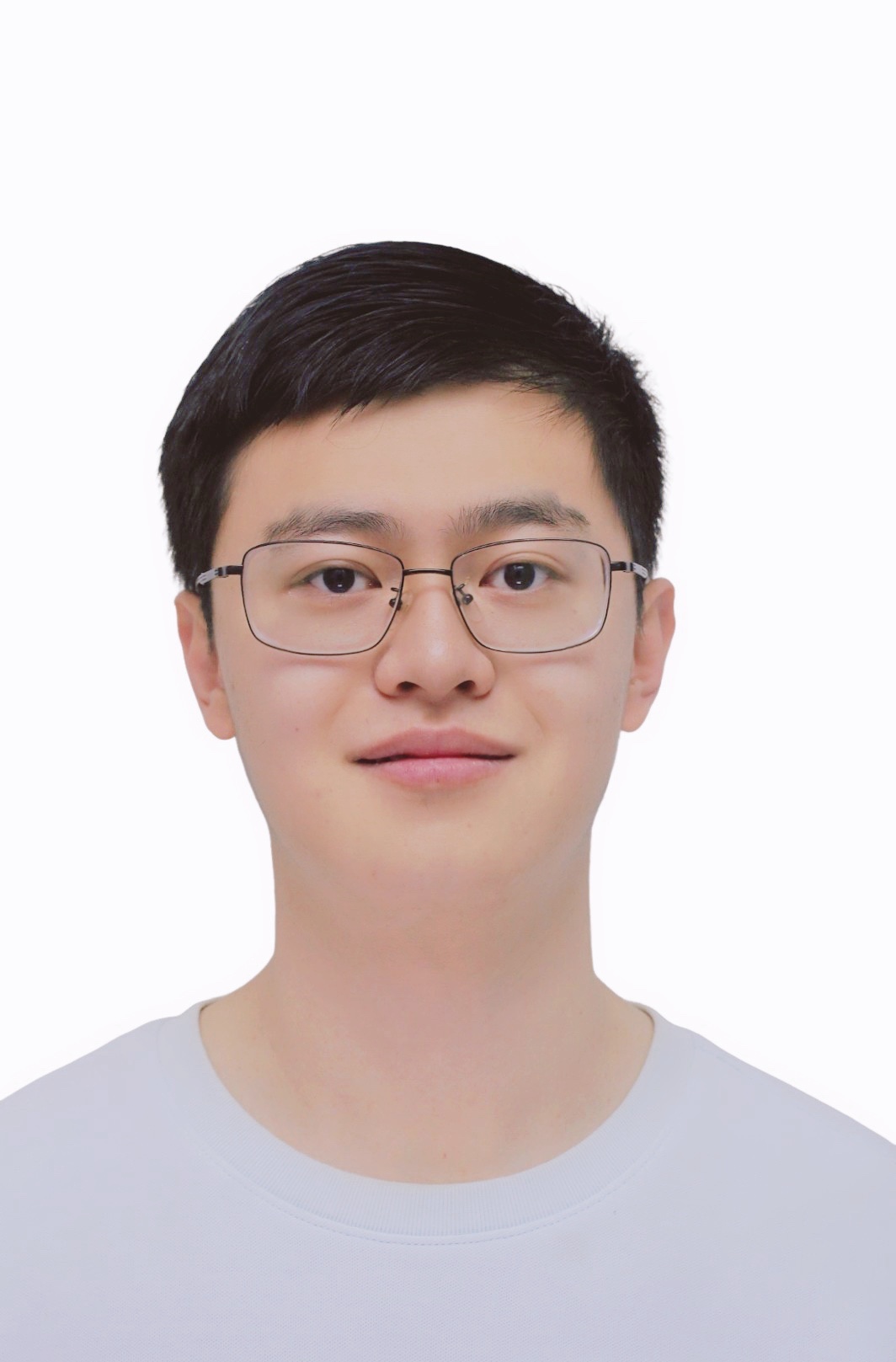}}]{Yinuo Deng} received his M.S. degree in 2025 from the College of Computer Science and Technology, Zhejiang University, Hangzhou, China and B.S. degree in 2022 from the School of Artificial Intelligence, Beijing University of Posts and Telecommunications, Beijing, China. He is currently with Alibaba Cloud, Hangzhou, China. His research interests include cloud computing, networking, and distributed systems.
\end{IEEEbiography}

\vspace{-1.3cm}

\begin{IEEEbiography}[{\includegraphics[width=1in,height=1.25in,clip,keepaspectratio]{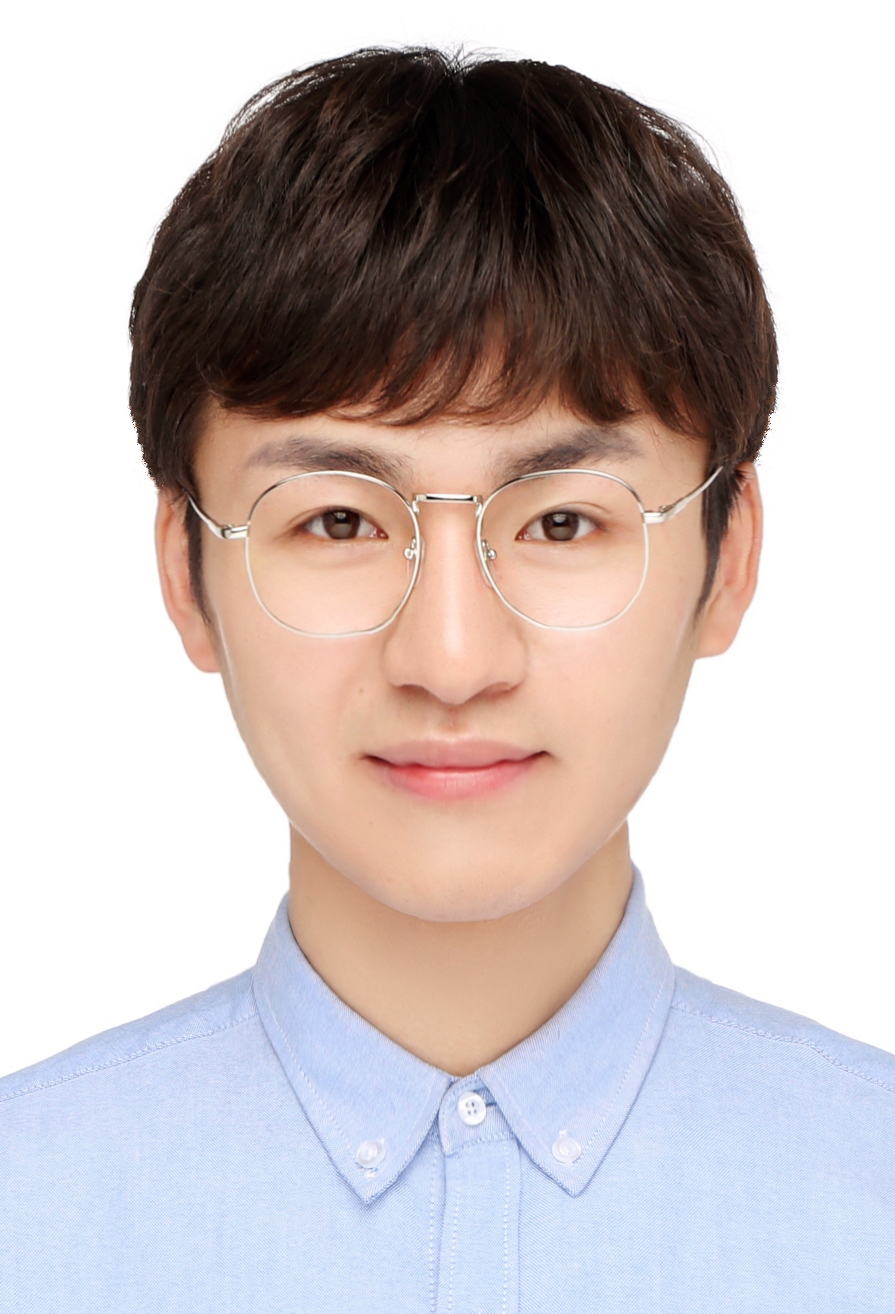}}]{Hailiang Zhao} is a ZJU 100 Young Professor at the School of Software Technology, Zhejiang University, and an Outstanding Qizhen Young Scholar. He received his Ph.D. in Computer Science from Zhejiang University in 2024, with a visiting research appointment at Nanyang Technological University, Singapore (2022-2023). His research lies at the intersection of services computing and service system performance optimization, with a focus on developing intelligent, learning-augmented algorithms and systems. He has authored or co-authored over 40 papers in leading journals and conferences, including Proceedings of the IEEE, IEEE TPDS, IEEE TMC, IEEE TSC, NeurIPS, CVPR, and ICWS. He serves as a regular reviewer for prestigious venues such as IEEE TSC, TKDD, Chinese Journal of Computers, FGCS, NeurIPS, and CVPR. He has received several honors, including the Incentive Program for Outstanding Ph.D. Dissertations by CCF-TCSC (2025), the Zhejiang University Outstanding Doctoral Dissertation Award (2024), and the Best Student Paper Award at IEEE ICWS 2019.
\end{IEEEbiography}

\vspace{-1.3cm}

\begin{IEEEbiography}[{\includegraphics[width=1in,height=1.25in,clip,keepaspectratio]{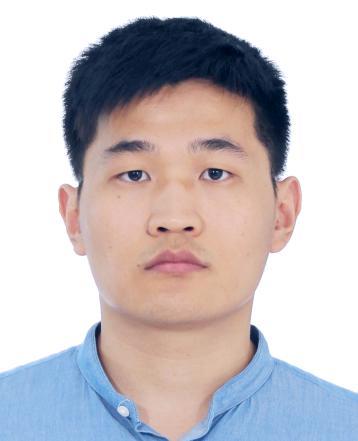}}]{Dongjing Wang}
		(Member, IEEE) received the B.S. and Ph.D. degrees in computer science from Zhejiang University, Hangzhou, China, in 2012 and 2018, respectively. He was cotrained at the University of Technology Sydney, Australia, for one year. He is currently a associate professor at Hangzhou Dianzi University, Hangzhou. He has published over 70 journal articles, including TMM, TPDS, TSC, TCYB, TNNLS, TOIS, and refereed conferences. His research interests include recommender systems, machine learning, data mining, and business process management.
	\end{IEEEbiography}

\vspace{-1.3cm}

\begin{IEEEbiography}[{\includegraphics[width=1in,height=1.25in,clip,keepaspectratio]{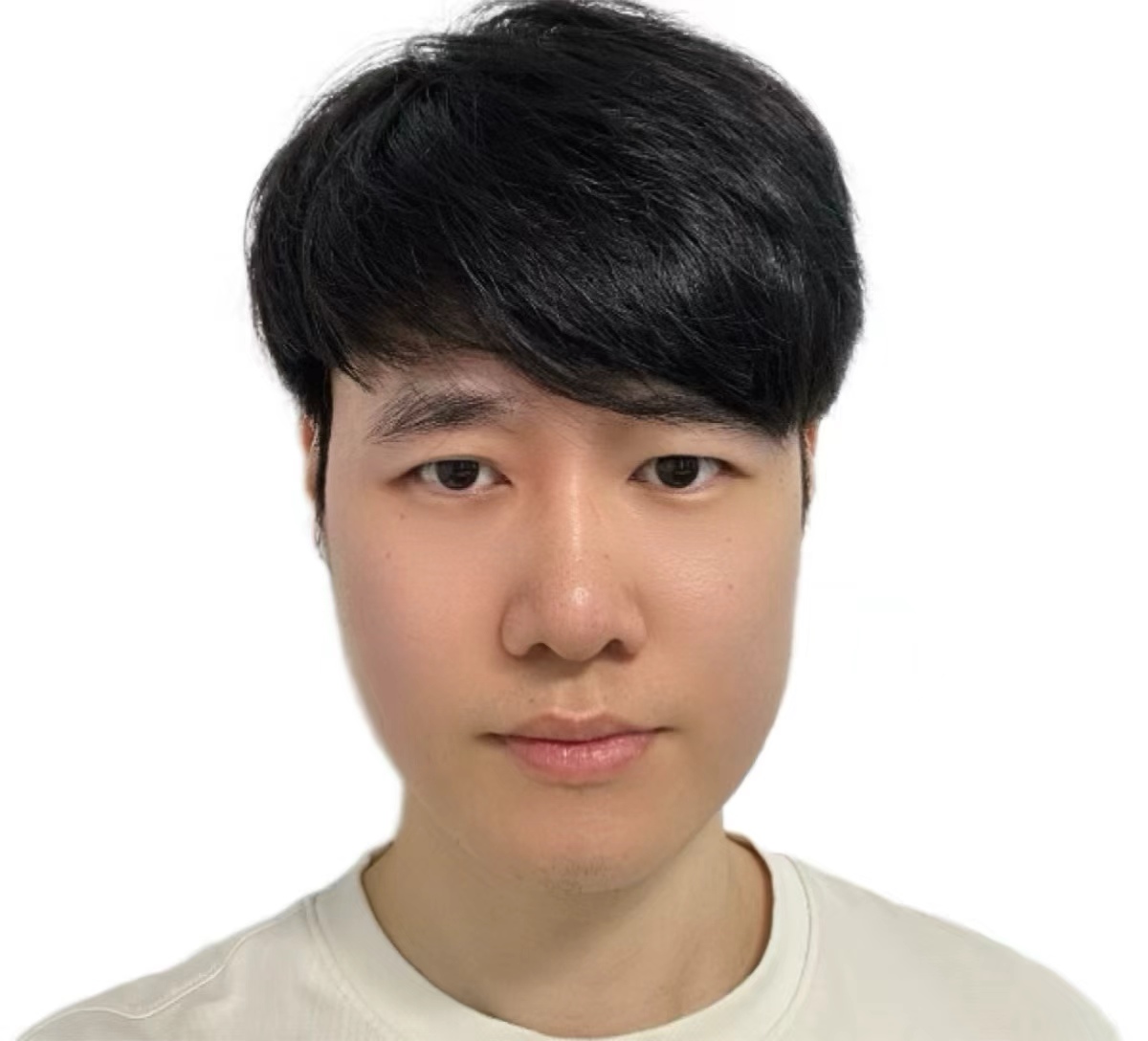}}]{Peng Chen} received the B.S. degree in Computer Science from Soochow University, China, and the M.S. degree in Informatics from Kyoto University, Japan, in 2020. He is currently pursuing the Ph.D. degree with the College of Computer Science and Technology, Zhejiang University, China. His research interests include theoretical machine learning, distributed systems, and service computing.
	\end{IEEEbiography}

\vspace{-1.3cm}

\begin{IEEEbiography}[{\includegraphics[height=1in,clip,keepaspectratio]{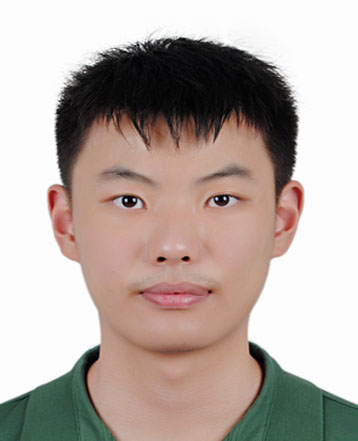}}]{Wenzhuo Qian} received the BS degree from the School of Computer Science and Technology, Hangzhou Dianzi University, Hangzhou, China, in 2023. He is currently working toward the Phd's degree in the College of Computer Science and Technology, Zhejiang University, Hangzhou, China. His research interests include edge computing and service computing.
	\end{IEEEbiography}

\vspace{-5mm}

\begin{IEEEbiography}
  [{\includegraphics[width=1in,height=1.25in,clip,keepaspectratio]{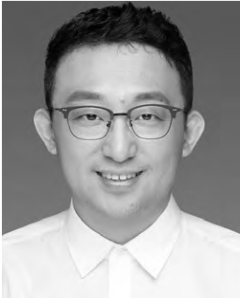}}]{Jianwei Yin} 
  received the Ph.D. degree in computer science from Zhejiang University (ZJU) in 2001. 
  He was a Visiting Scholar with the Georgia Institute of Technology. He is currently a Full Professor 
  with the College of Computer Science, ZJU. Up to now, he has published more than 100 papers in top 
  international journals and conferences. His current research interests include service computing 
  and business process management. He is an Associate Editor of the IEEE Transactions on Services 
  Computing.
\end{IEEEbiography}

\vspace{-0.5cm}

\begin{IEEEbiography}
  [{\includegraphics[width=1in,height=1.25in,clip,keepaspectratio]{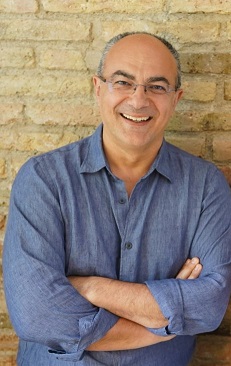}}]{Schahram Dustdar}
  is a Full Professor of Computer Science (Informatics) with a focus on Internet Technologies heading the Distributed Systems Group at the TU Wien. He was founding co-Editor-in-Chief of ACM Transactions on Internet of Things (ACM TIoT). He is Editor-in-Chief of Computing (Springer). He is an Associate Editor of IEEE Transactions on Services Computing, IEEE Transactions on Cloud Computing, ACM Computing Surveys, ACM Transactions on the Web, and ACM Transactions on Internet Technology, as well as on the editorial board of IEEE Internet Computing and IEEE Computer. Dustdar is recipient of multiple awards: TCI Distinguished Service Award (2021), IEEE TCSVC Outstanding Leadership Award (2018), IEEE TCSC Award for Excellence in Scalable Computing (2019), ACM Distinguished Scientist (2009), ACM Distinguished Speaker (2021), IBM Faculty Award (2012). He is an elected member of the Academia Europaea: The Academy of Europe, where the chairman of the Informatics Section for multiple years. He is an IEEE Fellow (2016), an Asia-Pacific Artificial Intelligence Association (AAIA) President (2021) and Fellow (2021). He is an EAI Fellow (2021) and an I2CICC Fellow (2021). He is a Member of the IEEE Computer Society Fellow Evaluating Committee (2022 and 2023).
\end{IEEEbiography}

\vspace{-0.5cm}

\begin{IEEEbiography}
  [{\includegraphics[width=1in,height=1.25in,clip,keepaspectratio]{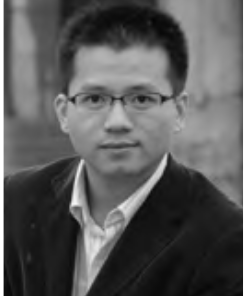}}]{Shuiguang Deng} 
  is currently a full professor at the College of Computer Science and Technology in Zhejiang University, China, 
  where he received a BS and PhD degree both in Computer Science in 2002 and 2007, respectively. He previously 
  worked at the Massachusetts Institute of Technology in 2014 and Stanford University in 2015 as a visiting scholar. 
  His research interests include Edge Computing, Service Computing, Cloud Computing, and Business Process Management. 
  He serves for the journal IEEE Trans. on Services Computing, Knowledge and Information Systems, Computing, and IET 
  Cyber-Physical Systems: Theory \& Applications as an Associate Editor. Up to now, he has published more than 100 
  papers in journals and refereed conferences. In 2018, he was granted the Rising Star Award by IEEE TCSVC. He is 
  a fellow of IET and a senior member of IEEE.
\end{IEEEbiography}

\end{document}